\documentclass[zpreprint,zbstnp]{zeus_paper}
\usepackage[english]{babel}
\usepackage{longtable}
\chardef\usc=95
\chardef\til=126
\catcode`\@=11 % @ signs are now treated as letters
\DeclareRobustCommand\xdotspace{\futurelet\@let@token\@xdotspace}
\def\@xdotspace{%
  \ifx\@let@token.\else
  \ifx\@let@token\bgroup.\else
  \ifx\@let@token\egroup.\else
  \ifx\@let@token\/.\else
  \ifx\@let@token\ .\else
  \ifx\@let@token~.\else
  \ifx\@let@token!.\else
  \ifx\@let@token,.\else
  \ifx\@let@token:.\else
  \ifx\@let@token;.\else
  \ifx\@let@token?.\else
  \ifx\@let@token/.\else
  \ifx\@let@token'.\else
  \ifx\@let@token).\else
  \ifx\@let@token-.\else
  \ifx\@let@token\@xobeysp.\else
  \ifx\@let@token\space.\else
  \ifx\@let@token\@sptoken.\else
   .\space
   \fi\fi\fi\fi\fi\fi\fi\fi\fi\fi\fi\fi\fi\fi\fi\fi\fi\fi}
\catcode`\@=12 % @ signs are no longer letters

\newcommand{\stru}[2]{%
   \relax\ifmmode\hbox{\vrule height#1 depth#2 width0pt}%
   \else\vrule height#1 depth#2 width0pt\fi}
\newcommand{\Ronum}[1]{\uppercase\expandafter{\romannumeral#1}}
\newcommand{\ronum}[1]{\expandafter{\romannumeral#1}}
\DeclareRobustCommand{\LaTeXZ}{%
  \LaTeX\kern-.05em4\kern-.1em
  {\raisebox{-0.2ex}{$\scriptstyle\text{ZEUS}$}}\xspace}
\DeclareMathAlphabet{\mathbf}{OT1}{cmr}{bx}{sl}
\newcommand{\eVdist}{\kern-0.06667em}

\newcommand{\slashfrac}[2]{%
  \raisebox{0.5ex}{\ensuremath #1}\kern-0.12em/\kern-0.08em
  \raisebox{-.8ex}{\ensuremath #2}}
\newcommand{\sqr}[3]{%
    {\vcenter{\hrule height.#3ex\hbox{\vrule width.#2ex height#1ex
     \kern#1ex\vrule width.#3ex}\hrule height.#2ex}}}

\catcode`\@=11 % @ signs are now treated as letters
\newcommand{\parenbar}{\mathpalette\p@renb@r}
\def\p@renb@r#1#2{\vbox{%
  \ifx#1\scriptscriptstyle \dimen@.7em\dimen@ii.2em\else
  \ifx#1\scriptstyle \dimen@.8em\dimen@ii.25em\else
  \dimen@1em\dimen@ii.4em\fi\fi \offinterlineskip
  \ialign{\hfill##\hfill\cr
    \vbox{\hrule width\dimen@ii}\cr
    \noalign{\vskip-.3ex}%
    \hbox to\dimen@{$\mathchar300\hfil\mathchar301$}\cr
    \noalign{\vskip-.3ex}%
    $#1#2$\cr}}}
\catcode`\@=12 % @ signs are no longer letters

\newcommand{\IP}{{\rm I$\kern-0.01667em$P}\xspace}

\newcommand{\F}{{\cal F}}

\mathchardef\qsm=63
\mathchardef\pls=43
\mathchardef\mns=512
\mathchardef\plm=518
\mathchardef\eql=61
\mathchardef\smallleft=300
\mathchardef\smallright=301
\mathchardef\les=316
\mathchardef\gre=318
\mathchardef\leq=532
\mathchardef\grq=533
\catcode`\@=11 % @ signs are now treated as letters
\newcounter{pict@width}
\newcounter{pict@height}
\newlength{\pict@scale}
\newcommand{\psfigadd}[4]{%
\setcounter{pict@width}{1*\ratio{#2+\pict@scale/2}{\pict@scale}}
\setcounter{pict@height}{1*\ratio{#3+\pict@scale/2}{\pict@scale}}
\setlength{\unitlength}{\pict@scale}
\hbox to #2{\hspace{-\fill}\begin{picture}(\thepict@width,\thepict@height)
\put(0,0){\psfig{figure=#1,width=#2,height=#3,clip=}}
\SetScale{0.283466457}
\SetWidth{1.763889}
{#4}
\end{picture}}
}
\newcounter{pict@widthfst}
\newcounter{pict@widthscd}
\newcounter{pict@widthtot}
\newcommand{\psfigaddtwo}[7]{%
\setcounter{pict@widthfst}{1*\ratio{#2+\pict@scale/2}{\pict@scale}}
\setcounter{pict@widthscd}{1*\ratio{#2+#4+\pict@scale/2}{\pict@scale}}
\setcounter{pict@widthtot}{1*\ratio{#2+#4+#6+\pict@scale/2}{\pict@scale}}
\setcounter{pict@height}{1*\ratio{#3+\pict@scale/2}{\pict@scale}}
\setlength{\unitlength}{\pict@scale}
\hbox{\hspace{-\fill}\begin{picture}(\thepict@widthtot,\thepict@height)
\put(0,0){\psfig{figure=#1,width=#2,height=#3,clip=}}
\put(\thepict@widthscd,0){\psfig{figure=#5,width=#6,height=#3,clip=}}
\SetScale{0.283466457}
\SetWidth{1.763889}
{#7}
\end{picture}}
}
\newcommand{\psfigror}[4]{%
\setcounter{pict@width}{1*\ratio{#2+\pict@scale/2}{\pict@scale}}
\setcounter{pict@height}{1*\ratio{#3+\pict@scale/2}{\pict@scale}}
\setlength{\unitlength}{\pict@scale}
\hbox{\begin{picture}(\thepict@width,\thepict@height)
\put(0,\thepict@height){\psfig{figure=#1,width=#3,height=#2,clip=,angle=270}}
\SetScale{0.283466457}
\SetWidth{1.763889}
{#4}
\end{picture}}
}
\newcommand{\psfigrol}[4]{%
\setcounter{pict@width}{1*\ratio{#2+\pict@scale/2}{\pict@scale}}
\setcounter{pict@height}{1*\ratio{#3+\pict@scale/2}{\pict@scale}}
\setlength{\unitlength}{\pict@scale}
\hbox{\begin{picture}(\thepict@width,\thepict@height)
\put(0,0){\psfig{figure=#1,width=#3,height=#2,clip=,angle=90}}
\SetScale{0.283466457}
\SetWidth{1.763889}
{#4}
\end{picture}}
}
\catcode`\@=12 % @ signs are no longer letters
\newlength\listtextwidth

\catcode`\@=11 % @ signs are now treated as letters
\newlength{\@tabfninsert}
\newlength{\@tabfnwidth}
\newcommand{\tabfootnote}[2]{%
  \setlength{\@tabfninsert}{0.8em}
  \setlength{\@tabfnwidth}{\textwidth}
  \addtolength{\@tabfnwidth}{-\@tabfninsert}
  \addtolength{\@tabfnwidth}{-0.4em}
  \noindent\makebox[\@tabfninsert][r]{\footnotesize$^{#1}$\hfil}\hfill%
  \parbox[t]{\@tabfnwidth}{\footnotesize #2\hfill}}
\catcode`\@=12 % @ signs are no longer letters

\def\etal{{\em et al.}}

\newcommand{\pom} {I\!\!P}
\newcommand{\reg} {I\!\!R}
\newcommand{\xpom}{x_{\xpom}}

\def\lsim{\mathrel{\rlap{\lower4pt\hbox{\hskip1pt$\sim$}}
    \raise1pt\hbox{$<$}}}         %less than or approx. symbol
\def\gsim{\mathrel{\rlap{\lower4pt\hbox{\hskip1pt$\sim$}}
    \raise1pt\hbox{$>$}}}         %greater than or approx. symbol

\def\Q2{\mbox{$Q^2$}}
\def\F2{\mbox{$F_2$}}

\newcommand\etjet{E_{T}^{\rm jet}}
\newcommand{\GeV}{{\rm GeV}}
\newcommand{\fitS}{ZEUS DPDF S}
\newcommand{\fitC}{ZEUS DPDF C}
\newcommand{\fitJ}{ZEUS DPDF SJ}

\def\citediff0{{\cite{%
zeusdiff0,*h1diff0%
}}\xspace}

\def\bpc1{{\cite{%
bpc01,*bpc02,*bpc03%
}}\xspace}

\begin{document}
%------------------------------------------------------------------------------
%       Title sheet
%------------------------------------------------------------------------------
\prepnum{{DESY-09-191}}

\title{
            A QCD analysis of ZEUS diffractive data
}                                                       
                    
\author{ZEUS Collaboration}
%\draftversion{1.0}
%\date{May 11th, 2009}
\draftversion{$3^{rd}$ EB}
%\date{.\ June 2004}
%\date{ }

\abstract{
%Diffractive parton distributions and their uncertainties 
%have been determined from a new next-to-leading order DGLAP QCD analysis 
%of ZEUS inclusive diffractive cross sections together with data on 
%diffractive dijet production. 
ZEUS inclusive diffractive cross-section measurements 
have been used in a DGLAP next-to-leading-order QCD 
analysis to extract the diffractive parton distribution functions. 
Data on diffractive dijet production in deep inelastic scattering 
have also been included to constrain the gluon density. 
%Diffractive photoproduction dijet data were used to
%test the extracted parton densities.
Predictions based on the extracted parton densities are   
compared to diffractive charm and dijet photoproduction data.  
}

\makezeustitle

\def\3{\ss}                                                                                        

\pagenumbering{Roman}                                                                              
                                    % this "%"s are for cosmetics only                             
                                                   %                                               
\begin{center}                                                                                     
{                      \Large  The ZEUS Collaboration              }                               
\end{center}                                                                                       
  S.~Chekanov,                                                                                     
  M.~Derrick,                                                                                      
  S.~Magill,                                                                                       
  B.~Musgrave,                                                                                     
  D.~Nicholass$^{   1}$,                                                                           
  \mbox{J.~Repond},                                                                                
  R.~Yoshida\\                                                                                     
 {\it Argonne National Laboratory, Argonne, Illinois 60439-4815, USA}~$^{n}$                       
\par \filbreak                                                                                     
  M.C.K.~Mattingly \\                                                                              
 {\it Andrews University, Berrien Springs, Michigan 49104-0380, USA}                               
\par \filbreak                                                                                     
  P.~Antonioli,                                                                                    
  G.~Bari,                                                                                         
  L.~Bellagamba,                                                                                   
  D.~Boscherini,                                                                                   
  A.~Bruni,                                                                                        
  G.~Bruni,                                                                                        
  F.~Cindolo,                                                                                      
  M.~Corradi,                                                                                      
\mbox{G.~Iacobucci},                                                                               
  A.~Margotti,                                                                                     
  R.~Nania,                                                                                        
  A.~Polini\\                                                                                      
  {\it INFN Bologna, Bologna, Italy}~$^{e}$
% ...
\newpage                                                                                           
$^{\    1}$ also affiliated with University College London,                                        
United Kingdom\\                                                                                   
% ...
%                                                                                                  
% \par         % if index listing & table fit to 1 page, put gap here                              
\newpage   % alternatively: go to newpage, if page is too small                                    
                                                           %                                       
% \institute_references_start    % do not touch or move this line !                                
                                                           %                                       
\begin{tabular}[h]{rp{14cm}}                                                                       
$^{a}$ &  supported by the Natural Sciences and Engineering Research Council of Canada (NSERC) \\  
$^{b}$ &  supported by the German Federal Ministry for Education and Research (BMBF), under        
          contract Nos. 05 HZ6PDA, 05 HZ6GUA, 05 HZ6VFA and 05 HZ4KHA\\                            
$^{c}$ &  supported in part by the MINERVA Gesellschaft f\"ur Forschung GmbH, the Israel Science   
          Foundation (grant No. 293/02-11.2) and the US-Israel Binational Science Foundation \\    
$^{d}$ &  supported by the Israel Science Foundation\\                                             
$^{e}$ &  supported by the Italian National Institute for Nuclear Physics (INFN) \\                
$^{f}$ &  supported by the Japanese Ministry of Education, Culture, Sports, Science and Technology 
          (MEXT) and its grants for Scientific Research\\                                          
$^{g}$ &  supported by the Korean Ministry of Education and Korea Science and Engineering          
          Foundation\\                                                                             
$^{h}$ &  supported by the Netherlands Foundation for Research on Matter (FOM)\\                   
$^{i}$ &  supported by the Polish State Committee for Scientific Research, project No.             
          DESY/256/2006 - 154/DES/2006/03\\                                                        
$^{j}$ &  partially supported by the German Federal Ministry for Education and Research (BMBF)\\   
$^{k}$ &  supported by RF Presidential grant N 1456.2008.2 for the leading                         
          scientific schools and by the Russian Ministry of Education and Science through its      
          grant for Scientific Research on High Energy Physics\\                                   
$^{l}$ &  supported by the Spanish Ministry of Education and Science through funds provided by     
          CICYT\\                                                                                  
$^{m}$ &  supported by the Science and Technology Facilities Council, UK\\                         
$^{n}$ &  supported by the US Department of Energy\\                                               
$^{o}$ &  supported by the US National Science Foundation. Any opinion,                            
findings and conclusions or recommendations expressed in this material                             
are those of the authors and do not necessarily reflect the views of the                           
National Science Foundation.\\                                                                     
$^{p}$ &  supported by the Polish Ministry of Science and Higher Education                         
as a scientific project (2009-2010)\\                                                              
$^{q}$ &  supported by FNRS and its associated funds (IISN and FRIA) and by an Inter-University    
          Attraction Poles Programme subsidised by the Belgian Federal Science Policy Office\\     
$^{r}$ &  supported by an FRGS grant from the Malaysian government\\                               
\end{tabular}                                                                                      
                                                           %                                       
% \institute_references_end     % do not touch or move this line !                                 
                                                           %                                       

%------------------------------------------------------------------------------
%       Text
%------------------------------------------------------------------------------
\pagenumbering{arabic} 
\pagestyle{plain}

% \def\wsstart{$\blacktriangleright\blacktriangleright\blacktriangleright$\par}
% \def\wsend{\par$\blacktriangleleft\blacktriangleleft\blacktriangleleft$}

% ----------------------------------------------------------------------------
%       Introduction
% ----------------------------------------------------------------------------

%================================================================
\section{Introduction}
\label{sec-int}

Many aspects of diffractive interactions can be described in the framework 
of quantum chromodynamics (QCD) as long as a hard scale is present, 
so that perturbative techniques can be used and the dynamics of the processes
can be formulated in terms of quarks and gluons. 
HERA data have contributed significantly to the understanding of such 
interactions, 
since events characterised by the diffractive dissociation of virtual photons, 
$\gamma^* p \rightarrow Xp$, 
constitute a large fraction ($\approx 10\%$) of the visible cross section in 
deep inelastic scattering (DIS).  
Diffractive reactions in DIS are a tool 
to investigate low-momentum partons in the proton, notably through the
study of diffractive parton distribution functions (DPDFs).
The latter are the densities of partons in the 
proton when the final state of the process contains a fast proton of
specified four-momentum. 
A precise knowledge of the DPDFs is an essential input to predictions 
of hard diffractive processes at the LHC. 

Several recent sets of DPDFs~\cite{h1-lrg, h1-dijets, MRW2007,GB2008}  
have been determined in global fits using the conventional 
DGLAP formalism~\cite{dglap1,dglap2,dglap3} in next-to-leading-order
(NLO) QCD. The diffractive structure function, $F_2^D$, which dominates 
the cross section, is directly sensitive to the quark density, whereas the 
gluon density is only indirectly constrained via scaling violations in the 
inclusive diffractive DIS cross sections. 
The inclusion of diffractive dijet data provides an additional 
constraint on the gluon density, since gluons directly contribute to jet 
production through the boson-gluon fusion process~\cite{Chekanov:2005nn}. 

For this paper, recently published ZEUS inclusive diffractive 
data~\cite{Chekanov:2008lrg} and diffractive DIS dijet 
data~\cite{Chekanov:2007dijetsdis} were used to extract the DPDFs. 
The inclusive data were first fitted alone and then 
in combination with the dijet data. The results are compared to H1 
fits~\cite{h1-lrg,h1-dijets}, as well as to ZEUS diffractive charm 
data~\cite{Chekanov:2003dstardis} and to ZEUS diffractive dijet 
photoproduction data~\cite{Chekanov:2007dijetsphp}.

%================================================================
\section{Theoretical framework}
\label{section:theory}

The cross section for diffractive DIS, $ep \to eXp$,  in the
one-photon-exchange approximation, can be expressed in terms 
of the diffractive reduced cross-section $\sigma_r^{D(3)}$: 
\begin{eqnarray}
\frac{d\sigma^{ep \rightarrow eXp}}{
d\beta dQ^2dx_{\pom}} & = &
\frac{2\pi\alpha^2}{\beta
  Q^4}\biggl[1+(1-y)^2\biggr]\sigma_r^{D(3)}(\beta,Q^2,x_{\pom}),
\label{sigma-2}
\end{eqnarray}
which depends on the diffractive structure functions, $F_{2/L}^{D(3)}$, as 

\begin{eqnarray}
\sigma_r^{D(3)}(\beta,Q^2,x_{\pom})
& = & 
F_2^{D(3)}(\beta,Q^2,x_{\pom}) - {y^2 \over 1 + (1-y)^2}\ F_{L}^{D(3)}(\beta,Q^2,x_{\pom}).
\label{sigma-2b}
\end{eqnarray}
%\noindent The quantity $R^D= \sigma_L^{\gamma^{\star} p \rightarrow 
%Xp}/\sigma_T^{\gamma^{\star} p \rightarrow Xp}$ is the ratio of the cross 
%sections for longitudinally and transversely polarised virtual photons. 
%The diiffractive longitudinal structure function, $F_L^D$, is related to 
%$R^D$ via $F_L^D=F_2^D R^D/(1+R^D)$. The diffractive reduced cross 
%section and the diffractive structure function coincide if $R^D=0$. 
%and $F_{\rm L}^{D(3)}$ is the diffractive longitudinal structure function.
The kinematic variables used in Eqs.~(\ref{sigma-2}) and (\ref{sigma-2b}) 
are defined as follows:   
\begin{itemize}
\item
$Q^2=-q^2=-(k-k')^2$, 
the negative invariant-mass squared of the exchanged virtual photon, 
where $q = k-k'$ is the difference 
of the four-momenta of the incoming and outgoing leptons; 
%Bjorken's $x=Q^2/(2P\cdot q)$, where 
%P is the four-momenta of the target proton; 
\item 
$x_{\pom}=(P-P')\cdot q/P\cdot q$, the fraction of the momentum of the 
proton carried by the diffractive exchange, where $P$ and $P'$ are 
the four-momenta of the incoming and outgoing protons, respectively; 
\item 
$\beta=Q^2/2(P-P')\cdot q$, the Bjorken variable defined for 
the diffractive exchange; 
 
\item 
the inelasticity $y=(q\cdot P)/(k \cdot P)$.  
\end{itemize}

The four-momentum exchanged at the proton vertex, $|t|$, is 
integrated over in Eqs.~(\ref{sigma-2}) and (\ref{sigma-2b}). 

The QCD factorisation theorem~\cite{qcdf,qcdf1,Trentadue,Berera} 
allows the diffractive structure functions, $F_{2/L}^{D(3)}$, 
to be expressed in terms of
convolutions of coefficient functions and DPDFs:

%, in
%the limit of large $Q^2$, at fixed $\beta$ and $\xpom$, 
%can be  written as

\begin{equation}
  \label{fact-theo}
F_{2/L}^{D(3)}(\beta,Q^2,x_{\pom}) =
 \sum_i \int_{\beta}^1 \frac{dz}{z}\,
     C_{2/L,i}\Big(\frac{\beta}{z}\Big)\, f_i^D(z,x_{\pom};Q^2) ,
\end{equation}
where 
the sum runs over partons of type $i$ and $z$ is the longitudinal 
momentum fraction of the parton entering the 
hard subprocess with respect to the diffractive exchange. 
In the lowest-order quark-parton model process, $z=\beta$. 
The inclusion of higher-order processes leads to $\beta<z$.  
The coefficient functions $C_{2/L,i}$ are the same as in inclusive DIS. 
In analogy to the usual parton distribution functions, the
DPDFs $f_i^D(z,x_{\pom};Q^2)$ are densities of partons of type 
$i$ with fractional momentum $z x_{\pom}$ in
a proton, probed with resolution $Q^2$ in a process with a fast proton in
the final state with fractional momentum ($1-x_{\pom}$).
The dependence of the DPDFs on the scale $Q^2$ is given by the DGLAP 
evolution equations.

%The factorisation theorem
%~(\ref{fact-theo}) 
%is applicable in the limit of large $Q^2$ and 
%fixed $x_{\pom}$; hence the DPDF extraction should be carried out  
%at each available $x_{\pom}$ value. 
%However, the range in $Q^2$ and $\beta$ is limited, depending on the value of 
%$x_{\pom}$, not allowing the full exploitation of the available statistics. 
%However, kinematic constraints make it 
%difficult to experimentally access the full range of $z$ and $Q^2$ 
%using data from only one value of
%$x_{\pom}$. Assumptions on the $x_{\pom}$ dependence of the DPDFs are 
%therefore necessary.  
Proton-vertex factorisation~\cite{regge} was adopted to model the $x_{\pom}$ 
dependence of the DPDFs. Two contributions were assumed, called 
Pomeron and Reggeon, 
separately factorisable into a term depending only on $x_{\pom}$ 
and a term depending only on $z$ and $Q^2$,

\begin{equation}
f_i^D(z,x_{\pom};Q^2)=
f_{\pom}(x_{\pom}) f_i(z,Q^2) + f_{\reg}(x_{\pom}) f_i^{\reg}(z,Q^2).
\label{i-s}
\end{equation}

The flux-factors $f_{\pom}$ and $f_{\reg}$ describe the emission of 
the Pomeron and Reggeon from the proton. Such an assumption was 
shown~\cite{Chekanov:2008lrg} to work to a good approximation 
for the data used in this analysis. 

%================================================================
\section{Analysis method}
\label{section:method}

The DGLAP evolution equations yield the distributions $f_{i}(z,Q^2)$ 
of the quarks and gluons at all values of $Q^2$, provided the DPDFs 
are parameterised as functions of $z$ at some starting scale $Q^2_0$. 
The input parameters were 
fitted to the data by minimising a $\chi^2$ function~\cite{Chekanov:2005nn}. 
Correlated systematic uncertainties were taken into account by using the 
method described in an earlier ZEUS publication~\cite{Chekanov:2002nlo}.

The QCD evolution was performed with the
programs {\sc Qcdnum}~\cite{f2bpc} as well as a newly developed package, 
{\sc Qcdc}\footnote{Computer code developed by W. Slominski.}.
The strong coupling constant was set to
$\alpha_s(M_Z)=0.118$. 
The contribution from heavy quarks was treated
within the general-mass variable-flavour-number scheme of Thorne and
Roberts (TR-VFNS)~\cite{tr2}, which interpolates between
the threshold and the high-$Q^2$ behaviour for heavy quarks.
%, as discussed in~\cite{Chekanov:2002nlo}. 
The values of the heavy-quark masses 
used were $m_c=1.35$ GeV and $m_b=4.3$ GeV~\cite{Chekanov:2005nn}. 
%the effect of varying these values is discussed in Section~\ref{section:sys}. 
The influence of $F_{L}^{D(3)}$ was accounted for through its 
NLO dependence on the parton densities. 

The NLO QCD predictions for the diffractive 
structure functions were obtained by 
convoluting the DPDFs with the QCD coefficient 
functions. 
%The latter include the dependence on the heavy quark
%masses according to the general-mass variable flavour-number 
%scheme of Thorne and Roberts (TR-VFNS)~\cite{tr2}.
Predictions for the inclusive diffractive reduced cross sections 
were obtained using $Q$ as the factorisation and renormalisation scales;
for the dijet cross sections, $Q$ was taken as the factorisation scale
and the transverse energy of the leading jet, $E_T^{\rm jet}$, as the
renormalisation scale. 
Predictions for the dijet cross sections at the parton level 
were performed at NLO with {\sc Disent}~\cite{disent} and 
{\sc Nlojet++}~\cite{nlojet}. 
The two programs agree within $5\%$.
These programs can deal with an arbitrary number of flavours but treat all
quarks as massless. Thus they match the TR-VFNS at scales much larger than
the quark masses, but may give imprecise results for scales 
close to the mass thresholds.
The predictions were corrected for hadronisation 
effects~\cite{tesi_bonato}.

%----------------------------------------------------------------
\subsection{Parameterisation of the DPDFs}
\label{section:input}

The DPDFs were modelled at the starting scale $Q_0^2=1.8$ GeV$^2$
in terms of quark singlet, $f_+ = \sum_{q}(f_q+f_{\bar q})$, 
and gluon, $f_g$, distributions. 
The neutrality of the diffractive exchange requires  
$f_{\bar q} = f_q$ for all flavours. 
The light-quark distributions were assumed to be equal, 
$ f_u = f_d = f_s $. 
The distributions of the $c$ and $b$ quarks 
were generated dynamically at the scale $Q$ 
above the corresponding mass threshold, i.e. no intrinsic charm 
or bottom were assumed. At the starting scale, chosen to be 
below the charm threshold, the quark-singlet 
parameterisation is summed over the light-quark distributions, 
$f_+ = 6 f_q$, where $q$ denotes any of $u,d,s$ and their antiquarks.
The distributions were parameterised at $Q_0^2$ as
\begin{eqnarray}
z f_{d,u,s}(z, Q_0^2) = A_{q}z^{B_{q}}(1-z)^{C_{q}},
\nonumber\\
z f_{g}(z, Q_0^2) = A_{g}z^{B_{g}}(1-z)^{C_{g}}.
\label{pdf_quark}
\end{eqnarray}

An additional factor, $e^{-\frac{0.001}{1-z}}$,
was included to ensure that the distributions vanish
for $z\rightarrow 1$ even for negative values of $C_{q,g}$.

The $x_{\pom}$ dependence was parameterised using Pomeron and 
Reggeon fluxes 

\begin{equation}
f_{\pom, \reg}(x_{\pom},t)=\frac{A_{\pom,\reg}e^{B_{\pom,
\reg}t}}{x_{\pom}^{2\alpha_{\pom,\reg}(t)-1}}~,
\label{flux}
\end{equation}
with linear trajectories $\alpha_{\pom, \reg}(t)=\alpha_{\pom, 
\reg}(0)+\alpha_{\pom, \reg}'t$, where $t$ is the four-momentum 
transfer  
at the proton vertex. The flux 
factors in Eq.~(\ref{flux}) were integrated 
over $t$ between $-1$~GeV$^2$ and the kinematically allowed 
maximum value, as for a previous ZEUS publication~\cite{Chekanov:2008lrg}. 
The values of the parameters which were
fixed in the fits are summarised in Table~\ref{tab-pomeronparam}.
The value of the normalisation parameter $A_{\pom}$ was 
absorbed in $A_{q,g}$. 
%As the relative normalization of the Pomeron flux 
%and of the DPDFs is arbitrary, the value of the 
%parameter $A_{\pom}$ was chosen such that 
%$x_{\pom} f_{\pom}(x_{\pom})=1$
% \cdot z \cdot f_i^D(z,Q^2)$ with 
%for $x_{\pom} = 0.003$. 
%where $f_{\pom(x_{\pom})}$.
The Reggeon parton densities, $f_i^{\reg}$ in
Eq.~(\ref{i-s}), were taken from a parameterisation derived from fits
to pion structure-function data~\cite{grv1}. 
%Modifying the parameters
%The effect of varying the fixed parameters within their errors was 
%negligible. 

In total, 9 parameters were left free in the fits: $A_{q,g}$,
$B_{q,g}$, $C_{q,g}$, the Pomeron and Reggeon intercepts, 
$\alpha_{\pom}(0)$ and $\alpha_{\reg}(0)$, 
and the normalisation of the Reggeon term, $A_{\reg}$. 

%----------------------------------------------------------------
\subsection{Data}
\label{section:datasets}

Inclusive diffractive data were fitted alone as well as in combination 
with a sample of diffractive dijet data. 
The inclusive diffractive reduced-cross-section data~\cite{Chekanov:2008lrg} 
used in the fits were selected using two methods: the requirement 
of a large rapidity gap between the final-state proton and the rest 
of the hadronic system (LRG sample) and the detection of the final-state 
proton in the ZEUS leading proton spectrometer 
(LPS sample)\footnote{The ZEUS measurements obtained with a third selection 
method based 
on the analysis of the distribution of the hadronic-final-state mass 
($M_X$ method)~\cite{mx1,mx2} were not included in the fits as this data 
sample is highly statistically correlated with the LRG sample.}. These data 
cover photon-proton centre-of-mass energies in the range $40<W<240\,\GeV$,
photon virtualities in the range 
$2<Q^2<305 \,\GeV^2$ (LRG)
and 
$2<Q^2<120 \,\GeV^2$ (LPS)
 and
hadronic-final-state masses in the range $2<M_X<25\,\GeV$
(LRG) and $2<M_X<40$\,GeV  (LPS). They 
span the following $x_{\pom}$ ranges: $0.0002<x_{\pom}<0.02$ (LRG) and
$0.002<x_{\pom}<0.1$ (LPS). Both samples are corrected for background 
from double-diffractive 
events $ep~\rightarrow~eXN$, in which the proton also 
dissociates into a low-mass state, $N$.  
The extracted DPDFs hence correspond to 
the single-diffractive reaction $ep~\rightarrow~eXp$. 
Only inclusive data with $Q^2 > 5$\,GeV$^2$ were included 
in the fits.
In order to minimise overlap between the two samples, 
LPS data with $x_{\pom} < 0.02 $ were excluded from the fit. 
Reduced cross sections in bins of 
$Q^2$, $x_{\pom}$ and $\beta$ were used.
%\footnote{As $t$ was not measured, Eq.(~\ref{fact-theo}) was 
%used integrated over $t$; as a consequence, 
%Eq.~(\ref{i-s}) did not depend on $t$.}.  
The total number of fitted data points was 229 from the LRG sample 
and 36 from the LPS sample.

The diffractive dijet sample in DIS~\cite{Chekanov:2007dijetsdis} used in 
the fits covers transverse energies $E_T^{\rm jet}>4$\,GeV and $x_{\pom}<0.03$. 
Differential cross sections as a function of $z_{\pom}^{\rm obs}$, an estimator of $z$, in different
$Q^2$ bins were used. The variable $z_{\pom}^{\rm obs}$ was calculated as 
$z_{\pom}^{\rm obs} = \frac{Q^2 + M_{jj}^2}{Q^2 + M_X^2}$, 
%\begin{equation}
%z_{\pom}^{\rm obs} = \frac{Q^2 + M_{jj}^2}{Q^2 + M_X^2}, 
%\nonumber
%\end{equation}
where $M_{jj}$ is the invariant mass of the dijet system. 
This sample provided 28 additional points.

The overlaps between the LRG and dijet samples and between the 
LPS and dijet samples are small. 
Therefore, the three data sets were considered statistically independent.   
The following sources of correlated systematic uncertainties 
were considered: 
\begin{itemize}
\item
LRG data: energy scale, reweighting of the simulation in $x_{\pom}$, variation 
of the energy threshold on the most forward energy-flow object and 
proton-dissociation background~\cite{Chekanov:2008lrg};
\item
LPS data: energy scale, reweighting of the simulation 
in $x_{\pom}$ and $t$ and proton-dissociation 
background~\cite{Chekanov:2008lrg};
\item
dijet data: energy scale and proton-dissociation 
background~\cite{Chekanov:2007dijetsdis}. 
\end{itemize}

The energy scale was taken to be common for all the data 
sets. The normalisation uncertainty 
due to the proton-dissociation background was taken as fully 
correlated between the LRG and dijet samples.

%An additional error 
%parameter was introduced to control\pagenumbering{arabic} 
\pagestyle{plain}

%================================================================
\section{Results}
\label{section:results}

Fits were performed to the data 
sets described in Section~\ref{section:datasets}.  
Two different parameterisations of the gluon density 
at the starting scale were used. This is summarised in Table~\ref{tab-fit}. 
% As the relative normalization 
% of the Pomeron flux, $f_{\pom}$, and of the DPDFs, $f_i(z,Q^2)$, in 
% Eq.~(\ref{i-s}) is arbitrary, the DPDFs are expressed in terms of 
% $x_{\pom} f_{\pom(x_{\pom})} \cdot z \cdot f_i(z,Q^2)$ with $x_{\pom} = 0.003$.

Two fits were performed to the LRG+LPS data: 
`Standard', with $A_g$, $B_g$ and $C_g$ 
as free parameters; `Constant', with $B_g=C_g =0$.
The corresponding sets of DPDFs are referred to as ``\fitS'' 
and ``\fitC'', respectively.

\begin{figure}[p]
\vfill
\begin{center}
\includegraphics[width=15cm,height=15cm]{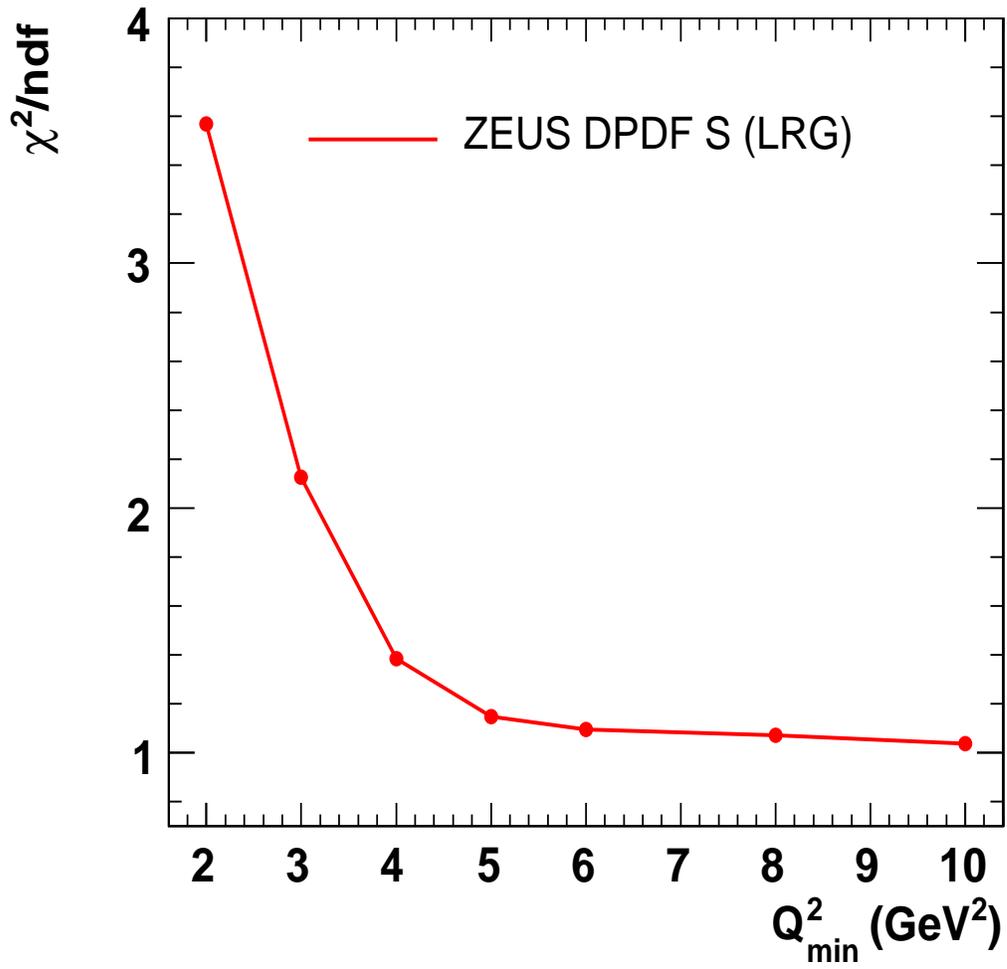}
\end{center}
\caption{Dependence of $\chi^2$/ndf for the fit ZEUS 
DPDF S on the value of the minimum $Q^2$, $Q_{\rm min}^2$, of the 
LRG data entering the fit.} 
\label{fig:chiQQmin}
\vfill
\end{figure}

Only data with $Q^2 > Q^2_{\rm min} =5 \,\GeV^2$
could be fitted within the combined assumptions of DGLAP evolution 
and proton-vertex factorisation.   
The quality of the fit drops rapidly for $Q^2_{\rm min} < 5 $\,GeV$^2$, as shown in  
Fig.~\ref{fig:chiQQmin}, where $\chi^2$/ndf obtained for the fit ZEUS 
DPDF~S, for statistical reasons 
restricted to the LRG data, is presented as a function of $Q^2_{\rm min}$.
Although the proton-vertex factorisation assumption worked reasonably well at low 
$Q^2$ for the LRG data~\cite{Chekanov:2008lrg}, the combination of this assumption 
with the DGLAP evolution in $Q^2$ breaks down below $Q^2$ of 5\,GeV$^2$. In contrast, 
in fully inclusive DIS, DGLAP fits were performed 
successfully~\cite{Chekanov:2002nlo} 
down to $Q^2$ of 2.5\,GeV$^2$. 

The results of the fits are listed in Table~\ref{tab-pdf}. 
In the minimisation procedure, the $\chi^2$ was calculated 
using only the statistical and uncorrelated systematic errors. 
The correlated systematic uncertainties were treated according to 
the ``offset method''~\cite{Chekanov:2002nlo} and were included in the 
determination of the full experimental uncertainty. To evaluate the 
goodness of the fit, the $\chi^2$ shown in Table~\ref{tab-pdf} was 
recalculated by adding 
the statistical, uncorrelated and correlated 
systematic errors in quadrature. The number of degrees of 
freedom in this case corresponds to the number of fitted points. 

The inclusive data are sensitive to all three quark-related parameters,  
$A_q$, $B_q$ and $C_q$. However, they show very little sensitivity to the 
gluon shape ($B_g$ and $C_g$). This follows from 
$F_{2}^{D(3)}$ being directly sensitive only to quarks. 
%All the parameters are well constrained by the fit 
%and 
The values obtained for $\alpha_{\pom}(0)$, $\alpha_{\reg}(0)$ 
and $A_{\reg}$ in all fits 
are fully consistent with those 
extracted from a fit~\cite{Chekanov:2008lrg} to the same data based 
on the proton-vertex factorisation assumption of Eq.~(\ref{i-s}). 

\begin{figure}[p]
\vfill
\begin{center}
\includegraphics[width=15cm,height=15cm]{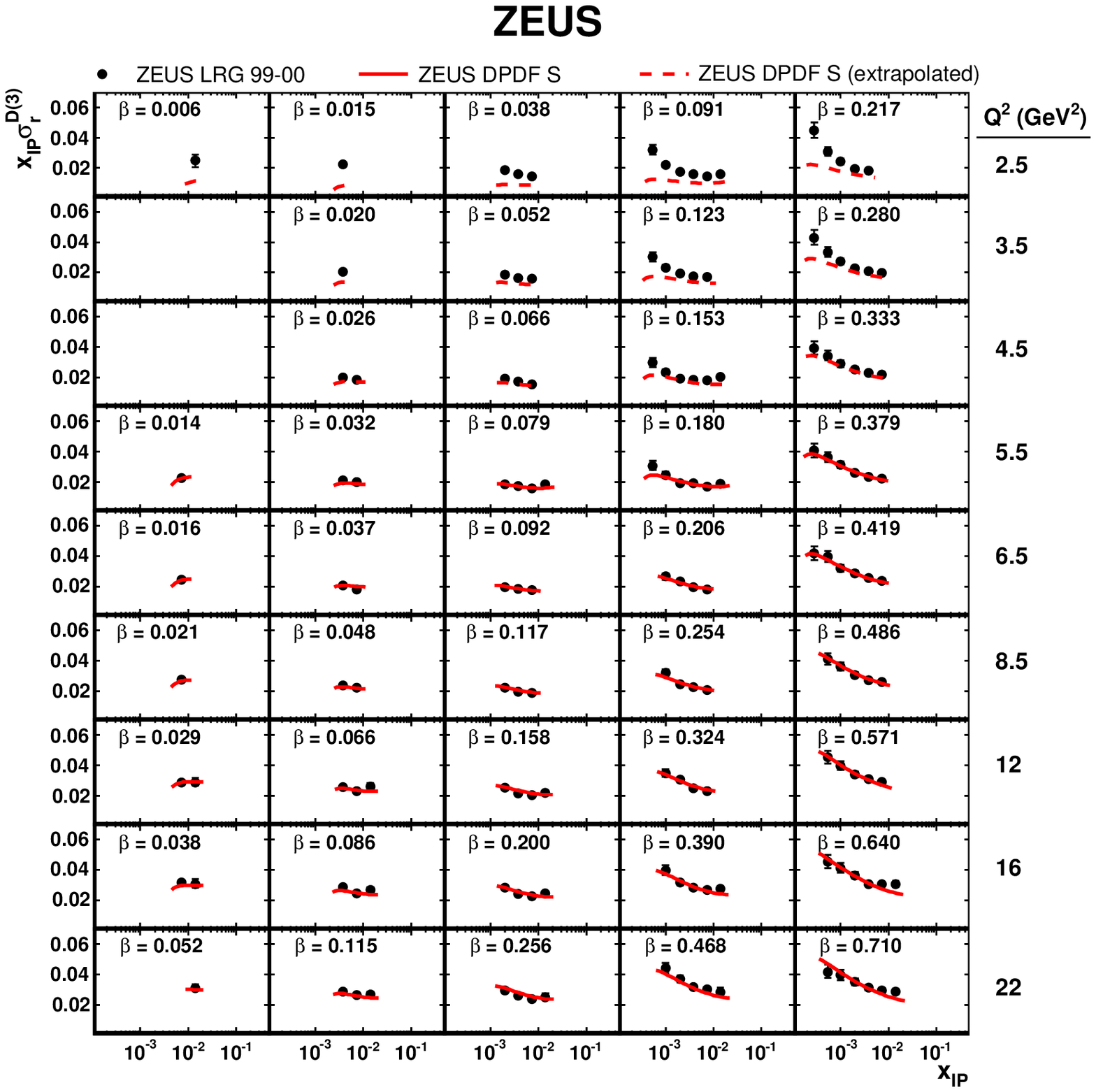}
\end{center}
%\vspace{3.0cm}
\caption{The fit \fitS\ compared to the ZEUS 
LRG data~\protect\cite{Chekanov:2008lrg} as a function of $x_{\pom}$ 
for different $\beta$ and $Q^2$ values at low $Q^2$. Where visible, 
the inner error bars show the statistical uncertainties and the full bars 
indicate the statistical and systematic uncertainties added in quadrature. 
The dashed lines represent the DGLAP extrapolation beyond the fitted region. 
}
\clearpage
\label{fig:LRG1}
\vfill
\end{figure}

\begin{figure}[p]
\vfill
\begin{center}
\includegraphics[width=15cm,height=15cm]{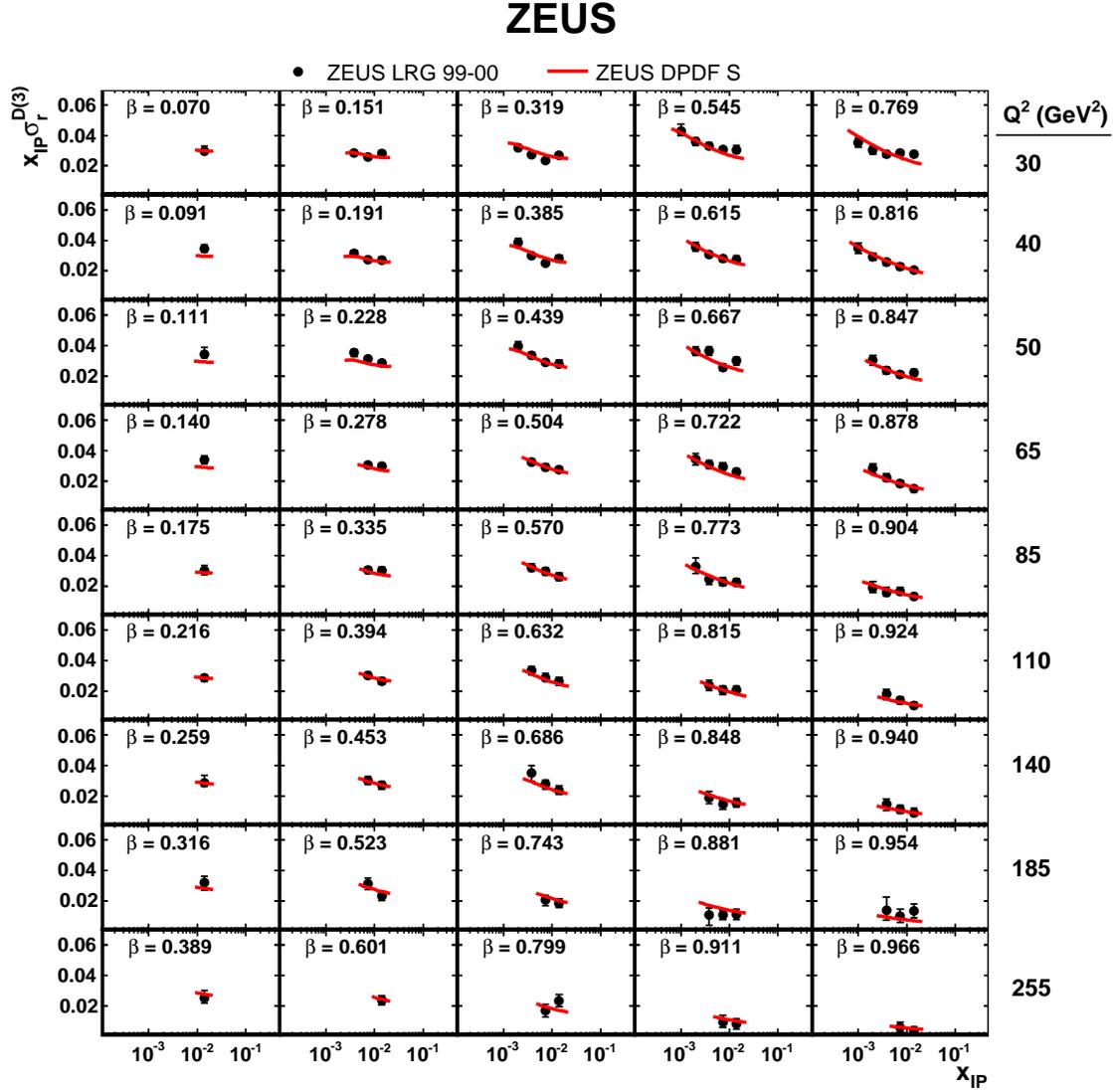}
\end{center}
%\vspace{3.0cm}
\caption{The fit \fitS\ compared to the ZEUS 
LRG data~\protect\cite{Chekanov:2008lrg} as a function of $x_{\pom}$ 
for different $\beta$ and $Q^2$ values at high $Q^2$. Where visible, 
the inner error bars show the statistical uncertainties and the full bars 
indicate the statistical and systematic uncertainties added in quadrature.
}
\clearpage
\label{fig:LRG2}
\vfill
\end{figure}

\begin{figure}[p]
\vfill
\begin{center}
\includegraphics[width=0.8\columnwidth]{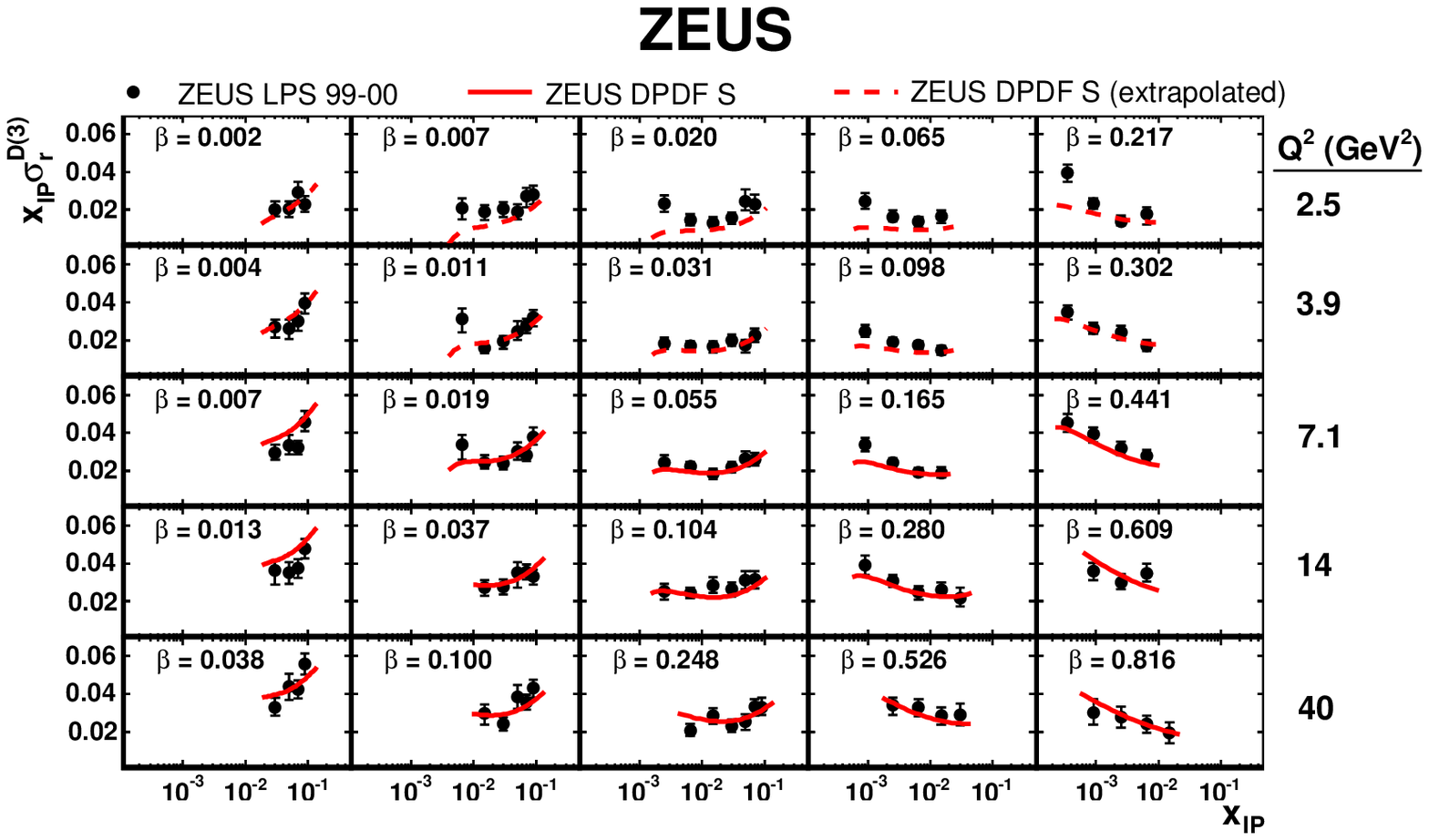}
\end{center}
%\vspace{3.0cm}
\caption{The fit \fitS\ compared to the ZEUS 
LPS data~\protect\cite{Chekanov:2008lrg} as a function of $x_{\pom}$ 
for different $\beta$ and $Q^2$ values. The inner error bars show 
the statistical uncertainties and the full bars indicate the 
statistical and systematic uncertainties added in quadrature. The dashed 
lines represent the DGLAP extrapolation beyond the fitted region. 
}
\clearpage
\label{fig:LPS}
\vfill
\end{figure}

Fits ZEUS DPDF C and S are of equally good quality and the 
predicted reduced cross sections are indistinguishable. 
Figures~\ref{fig:LRG1}--\ref{fig:LPS} show fit S 
compared to the LRG and LPS data. Both data samples are 
well described. 
For $Q^2 < 5\,\GeV^2$, the predictions are extrapolated and 
underestimate the LRG as well as the LPS data for $x_{\pom} < 0.005$. 
The fit is above the LPS data in the low-$\beta$ region, where there 
are no LRG data. 

\begin{figure}[p]
\vfill
\centerline{\includegraphics[width=0.6\columnwidth]{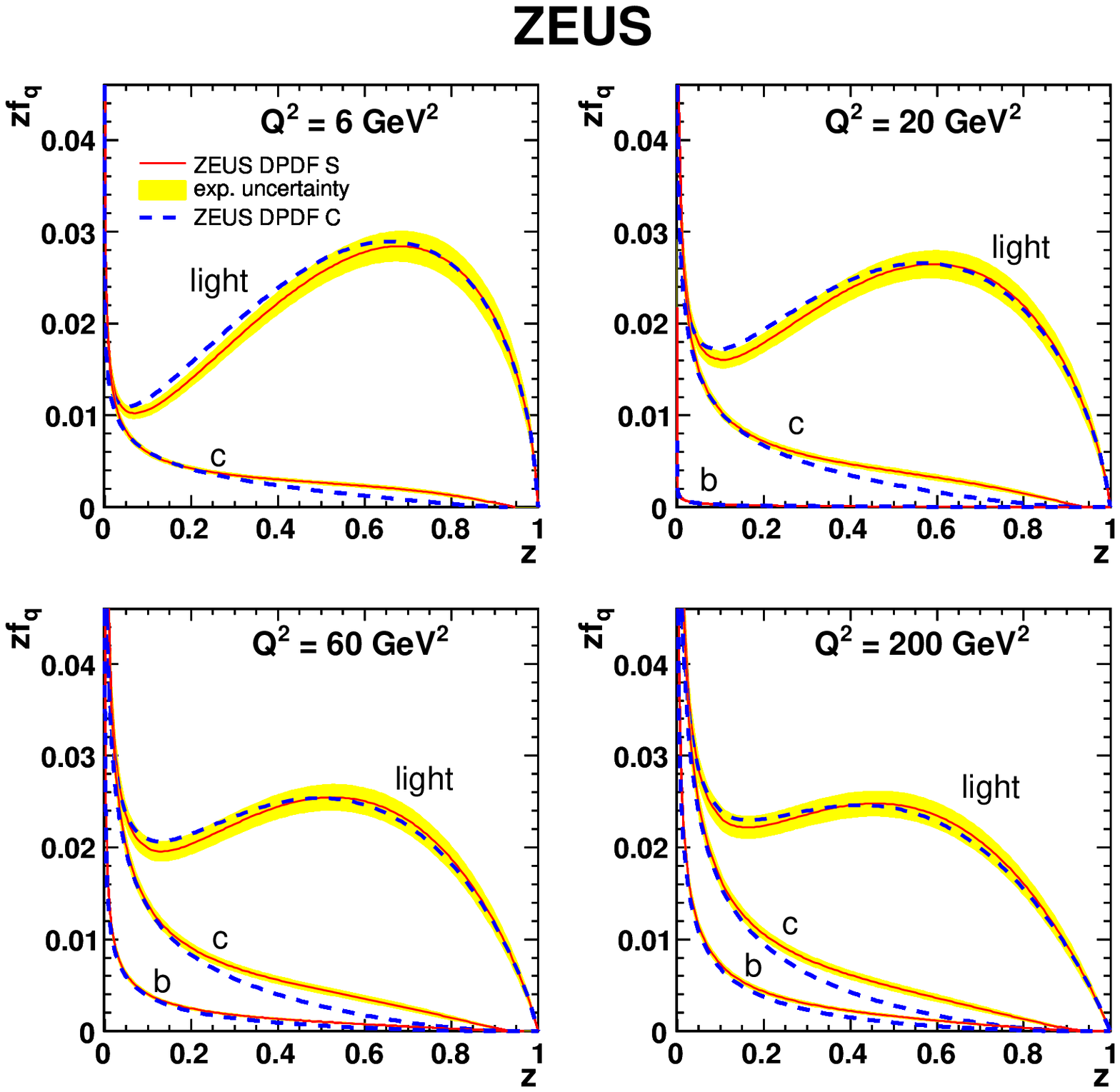}}
\vspace{5mm}
\centerline{\includegraphics*[width=0.6\columnwidth,bb= 0 0 482 440]{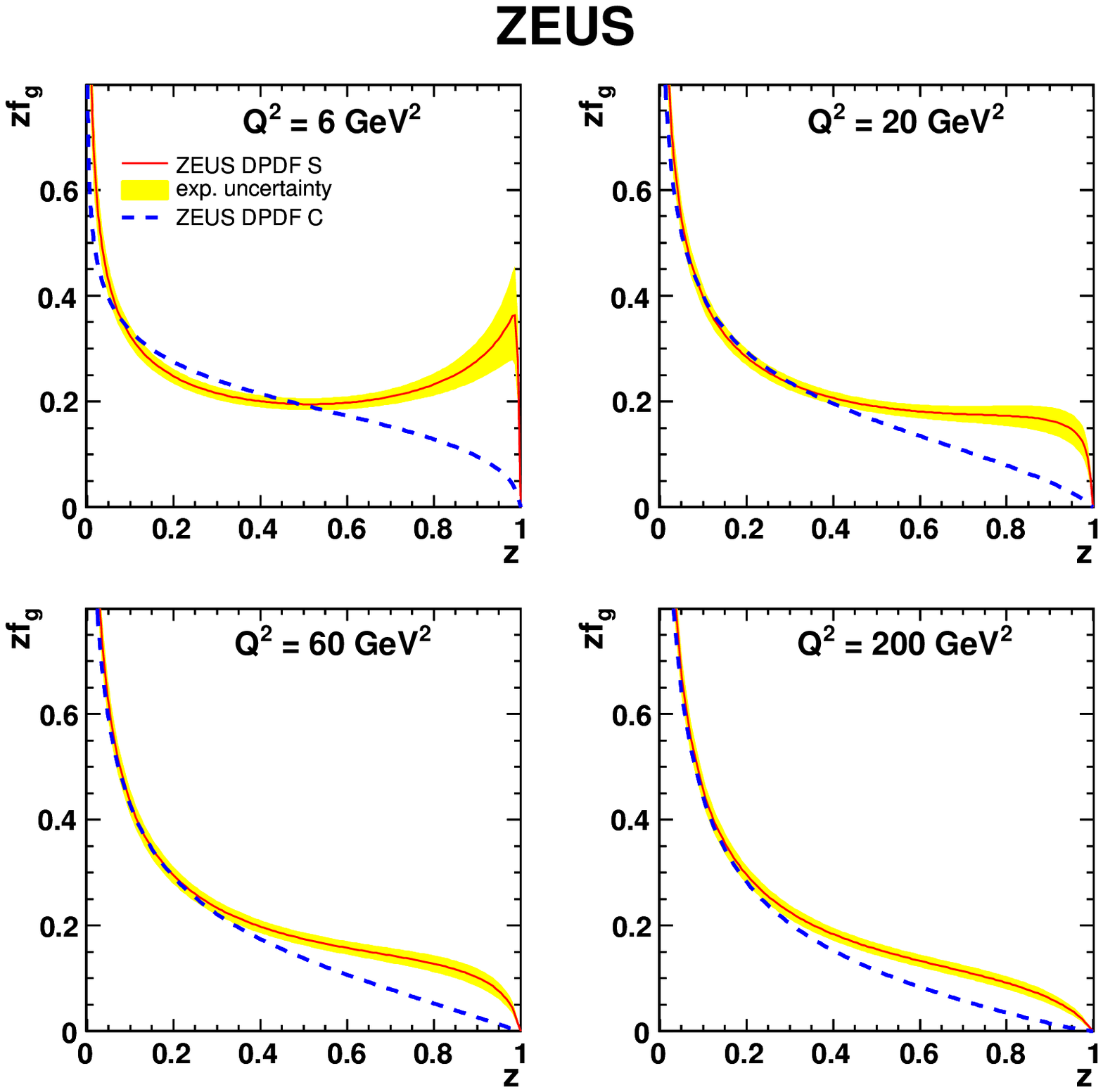}}
\caption{Four upper (lower) plots: the quark (gluon) 
distributions 
obtained from fits \fitS\ (continuous line) 
and \fitC\ (dashed line), 
shown for four different values of $Q^2$. 
%Four lower plots: 
%the gluon distribution obtained from fits ZEUS DPDF S-incl  
%(continuous line) and ZEUS DPDF C-incl (dashed line), 
%shown at 4 different values of $Q^2$. 
The shaded error bands show the experimental uncertainty.} 
%FIXME:WHAT INCLUDED?
%experimental correlated systematic uncertainties, normalisations 
%and model uncertainties
\label{fig:pdfsSC}
\vfill
\end{figure}

The quark and gluon densities, $zf_q(z,Q^2)$ and $zf_g(z,Q^2)$, from fit S
are shown with their experimental uncertainties in the upper (lower) part of 
Fig.~\ref{fig:pdfsSC} for $Q^2=6$, 20, 60 and 200 GeV$^2$. 
The DPDFs from fit C are also shown.
The relative normalisation 
of the Pomeron flux, $f_{\pom}$, and of the DPDFs, $f_i(z,Q^2)$, in 
Eq.~(\ref{i-s}) is arbitrary; for Fig.~\ref{fig:pdfsSC} the normalisation 
was chosen such that at $x_{\pom} = 0.003$ the quantity $x_{\pom} f_{\pom}(x_{\pom})=1$. 
%\footnote{This is equivalent to the flux normalisation
%$x_{\pom} f_{\pom}(x_{\pom})=1$ at $x_{\pom} = 0.003$, as in~\cite{h1-lrg}.}.
% ws -->

The quark distributions are very similar for the two fits
while the gluon densities are significantly different.
Gluons from fit S grow rapidly at high $z$, 
while those from fit C vanish as $z \rightarrow 1$ 
in a smoother way.
The large discrepancy demonstrates 
the low sensitivity of the inclusive data to gluons. 
To constrain the gluons better, a more exclusive process is 
needed where photon-gluon fusion contributes at leading order. 

\begin{figure}[p]
\vfill
\begin{center}
\includegraphics[width=1.\columnwidth]{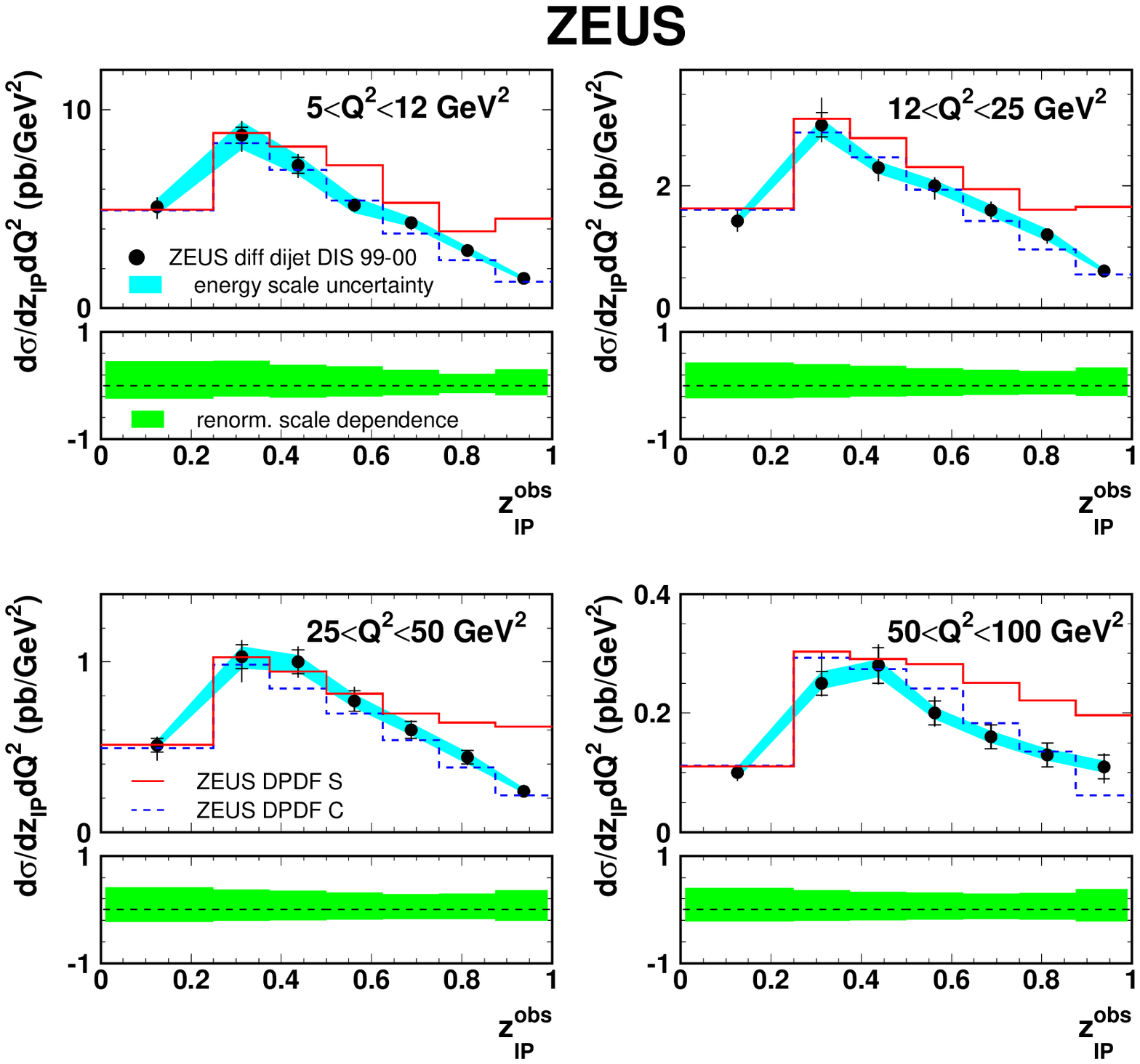}
\end{center}
\caption{Predictions based on \fitS\ (continuous line) 
and \fitC\ (dashed line) compared to the ZEUS diffractive 
dijet data~\protect\cite{Chekanov:2007dijetsdis} 
as a function of $z_{\pom}^{\rm obs}$ for different $Q^2$ values. 
The inner error bars show 
the statistical uncertainties and the full bars indicate the 
statistical and systematic uncertainties added in quadrature.
The dark shaded bands indicate the jet energy scale 
uncertainty. 
%Correlated uncertainty among the data points. 
%The light 
%shaded bands show the uncertainty due to the variation 
%of the renormalisation scale.    
%The uncertainty due to the variation of the renormalisation 
%scale, of the order of 20-40\%, is not shown. 
The light shaded bands at the bottom of each plot show the renormalisation 
scale uncertainty. 
}
\label{fig:dj-CS}
\vfill
\end{figure}

Predictions based on \fitS\ and C are compared 
to the double-differential diffractive dijet cross 
section~\cite{Chekanov:2007dijetsdis} in 
Fig.~\ref{fig:dj-CS}. The uncertainty due to the 
variation of the renormalisation scale in the 
calculation between 0.5$E_T^{\rm jet}$ and 
2$E_T^{\rm jet}$ is also shown. 
At high $z_{\pom}^{\rm obs}$ 
the predictions based on fit C give a good description of the dijet 
data throughout the whole kinematic region.
% a compelling demonstration of QCD factorisation. 
This is not the case for fit S. Figure~\ref{fig:dj-CS} 
clearly shows that the dijet data are sensitive to the 
gluon density. 
%and can be used to constrain it. 

\begin{figure}[p]
\vfill
\begin{center}
%%%\framebox[13.cm]{\rule{0.pt}{10.cm}}
\includegraphics[width=1.\columnwidth]{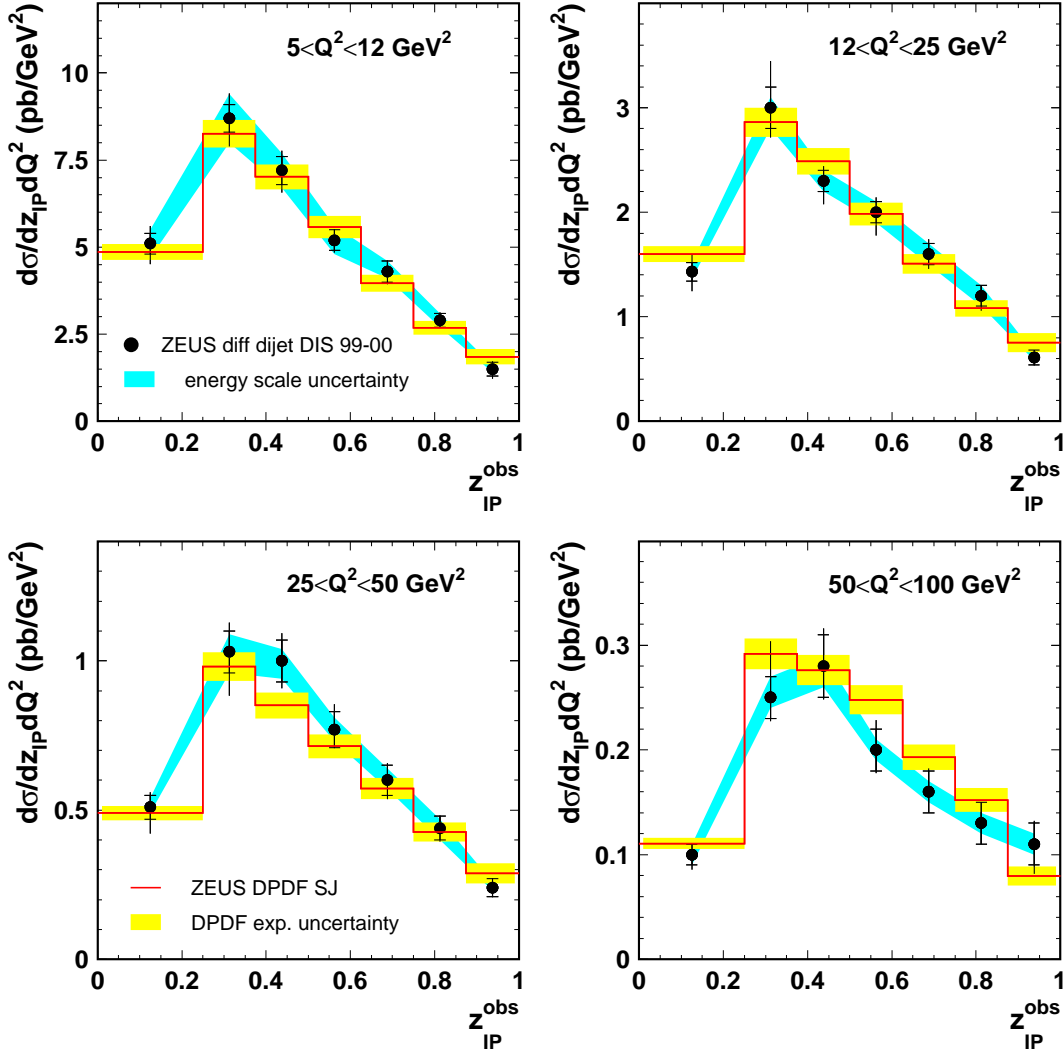}
\end{center}
\caption{The fit \fitJ\ compared to the 
ZEUS diffractive 
dijet data~\protect\cite{Chekanov:2007dijetsdis} 
as a function of $z_{\pom}^{\rm obs}$ for different $Q^2$ values. 
The inner error bars show 
the statistical uncertainties and the full bars indicate the 
statistical and systematic uncertainties added in quadrature.
The dark shaded bands indicate the jet energy scale 
uncertainty. 
The light shaded bands show the DPDF experimental uncertainties. 
}
\label{fig:dj-J}
\vfill
\end{figure}

A third fit was performed to the LRG+LPS data in combination with 
the dijet data. The corresponding DPDFs are referred to as 
``\fitJ'' and the resulting parameters are also given in 
Table~\ref{tab-pdf}. This fit is indistinguishable from fit S 
when compared to the inclusive data and also provides 
a remarkably good description of the dijet data, as shown 
in Fig.~\ref{fig:dj-J}.
% again a proof of QCD factorisation. 

\begin{figure}[p]
\vfill
\centerline{\includegraphics[width=0.6\columnwidth]{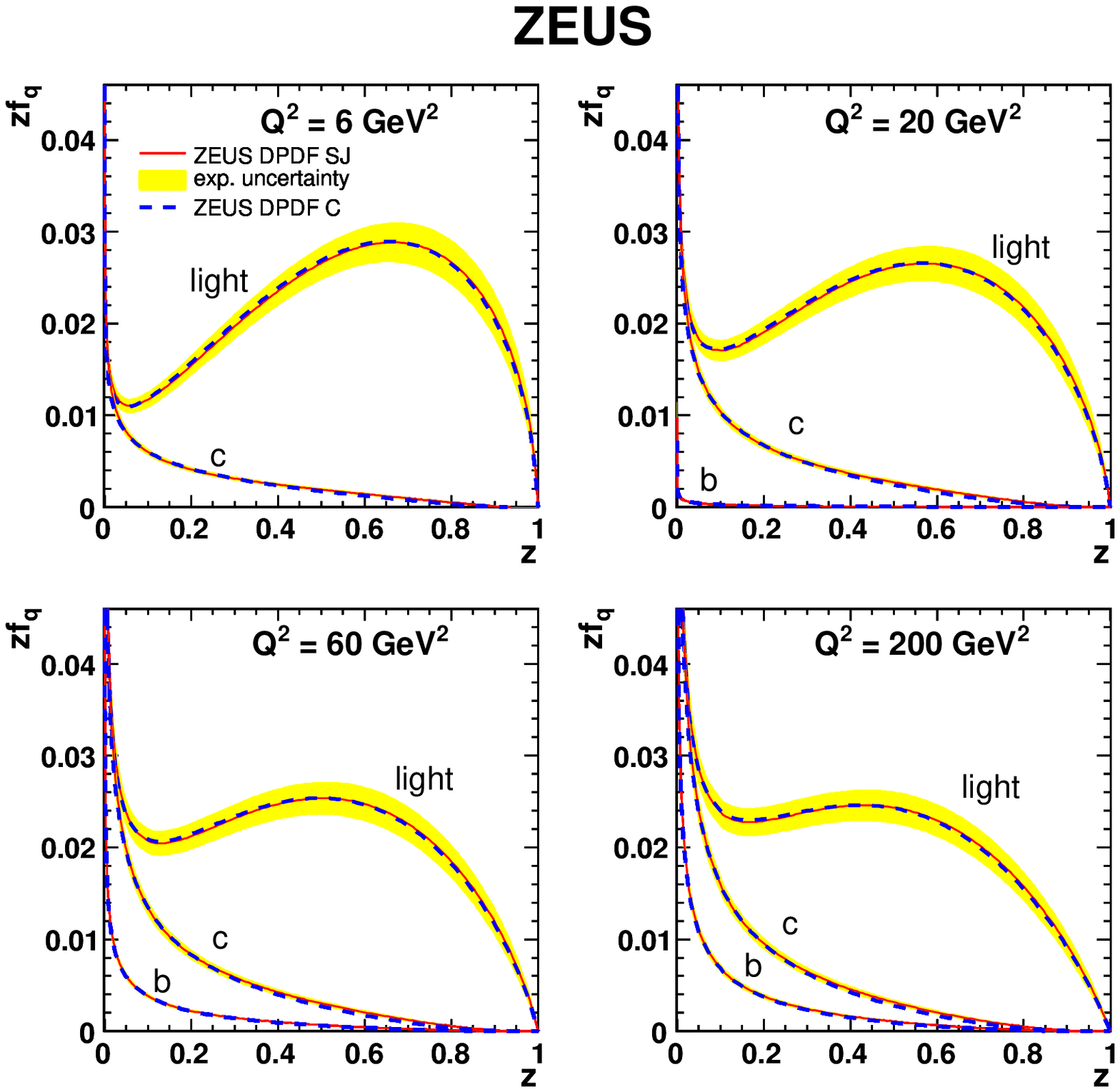}}
\vspace{5mm}
\centerline{\includegraphics*[width=0.6\columnwidth,bb= 0 0 482 440]{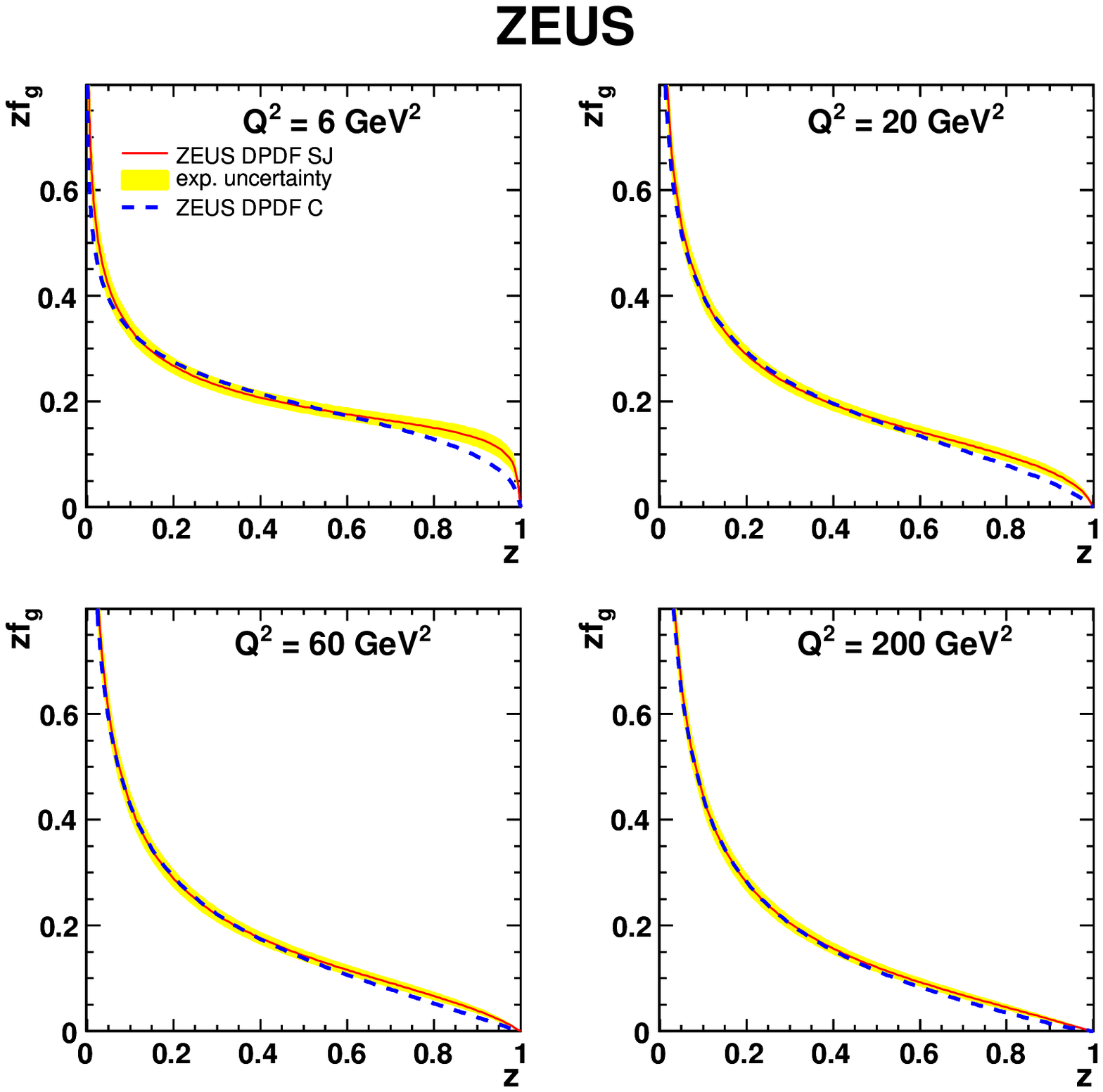}}
\caption{Four upper (lower) plots: the quark (gluon) 
distributions obtained from 
fits \fitJ\ (continuous line) and \fitC\ (dashed line), 
shown for four different values of $Q^2$. 
%For lower plots: the gluon distribution obtained from fits ZEUS DPDF 
%S-incl+dijets (continuous line) and ZEUS DPDF C-incl (dashed line), 
%shown at 4 different values of $Q^2$. 
The shaded error bands show the experimental uncertainty.} 
\label{fig:pdfsJ}
\vfill
\end{figure}

The quark and gluon densities from fit SJ 
are shown with their experimental uncertainties in Fig.~\ref{fig:pdfsJ} 
for $Q^2=6$, 20, 60 and 200 GeV$^2$. The result of fit C is also shown 
for comparison.
Again, the plotted quantities are $zf_{q,g}(z,Q^2)$ with the normalisation 
$x_{\pom} f_{\pom}(x_{\pom}) = 1 $ at $x_{\pom} = 0.003$. 
The decrease in the uncertainty on the gluon distribution with respect 
to the fit without jet data (Fig.~\ref{fig:pdfsSC}) is clearly seen. 
Combining the inclusive and dijet data constrains the gluon and the
quark densities with a comparable precision across the whole $z$ range.

\begin{figure}[p]
\vfill
\begin{center}
\includegraphics[width=0.8\columnwidth]{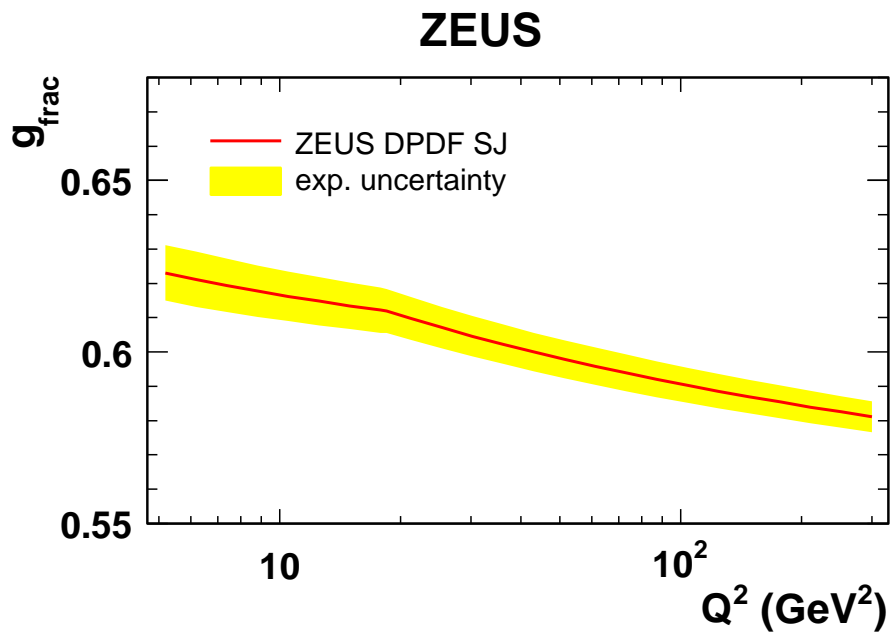}
\end{center}
\caption{$Q^2$ dependence of the gluon momentum fraction, $g_{\rm frac}$,  
according to the fit \fitJ. The shaded error band 
shows the experimental uncertainty.} 
\label{fig:glu}
\vfill
\end{figure}

Figure~\ref{fig:glu} shows the $Q^2$ dependence of the 
fraction of the longitudinal momentum of the diffractive 
exchange carried by the gluons, $g_{\rm frac}$, according to fit
\fitJ, integrated over the range $10^{-5} < z < 1$.
Over the wide range $5 < Q^2 < 300$ GeV$^2$, the fraction 
amounts to approximately 60\%. 
% <-- ws 2009-05-25
The fall with $Q^2$ is a direct consequence of the DGLAP evolution which forces
$g_{\rm frac}$ to approach $\approx 0.5$  
at high $Q^2$, while the slope change at $Q = m_b$ 
reflects the change in the number of flavours from four to five.

%----------------------------------------------------------------
\subsection{DPDF uncertainties}
\label{section:sys}

%FIXME:$Q_0^2$ and 
%VARIATION OF FIXED PARAMETERS CAN BE MENTIONED HERE INSTEAD 
%OF in 3.2

The following sources of uncertainties were investigated: 

\begin{itemize}
\item 
the starting scale $Q^2_0$. The value $1.8$ GeV$^2$ was chosen as 
it minimises the $\chi^2$. 
%The input scale was set to $Q_0^2=1.8 GeV^2$ 
Varying $Q_0^2$ for fit C between 1.6 and 2 GeV$^2$ yielded a $\chi^2$ 
between 1.18 and 1.20; the DPDFs did not change significantly;

\item 
the fixed parameters in the fits (Table~\ref{tab-pomeronparam}). 
Variations within the measurement errors resulted 
in a simple scaling of the fluxes integrated over $t$, absorbed into 
the normalisation parameters $A_{q}$, $A_{g}$ and $A_{\reg}$, with 
negligible effect on the DPDFs;

\item 
the renormalisation scale dependence.  
The scale $\mu_R$ for dijet data was taken as 
0.5$E_T^{\rm jet}$ and 2$E_T^{\rm jet}$, 
whereas it was kept as $Q$ for the inclusive data. 
The effect on the parton densities was within 5\% for light quarks, 
15\% for $c$ and $b$ and 30\% for gluons, while the $\chi^2$ 
increased significantly;

\item the masses of the charm and beauty quarks. 
The nominal values of $m_c=1.35\,\GeV$ and 
$m_b=4.3\,\GeV$ were varied in the ranges $1.35 < m_c < 1.75\,\GeV$
and $4.3 < m_b < 5\,\GeV$.
%  and the corresponding effect on the DPDFs is shown in 
% Fig.~\ref{fig11}; 
Neither the quark nor gluon distributions were 
sensitive to variations of $m_b$, whereas $m_c$ produced an 
effect comparable to the experimental uncertainty.
The $\chi^2$ value changed only slightly,  
reaching the minimum at the nominal mass values.

\end{itemize}

These theoretical uncertainties are not shown in 
Figs.~\ref{fig:pdfsSC},~\ref{fig:pdfsJ} and ~\ref{fig:glu}. Given are the 
total experimental uncertainties as determined with the offset 
method~\cite{Chekanov:2002nlo}.

%----------------------------------------------------------------
\subsection{Comparison to other HERA DPDFs}

An earlier ZEUS DPDF fit to inclusive diffractive measurements 
combined with data on charm production~\cite{lps+charm} 
did not include data points with values of $\beta>0.4$.
Also, in the current analysis a far larger data sample is used. 
This previous analysis is therefore superseded by the results of 
the present paper.  

\begin{figure}[p]
\vfill
\begin{center}
\includegraphics[width=15cm,height=15cm]{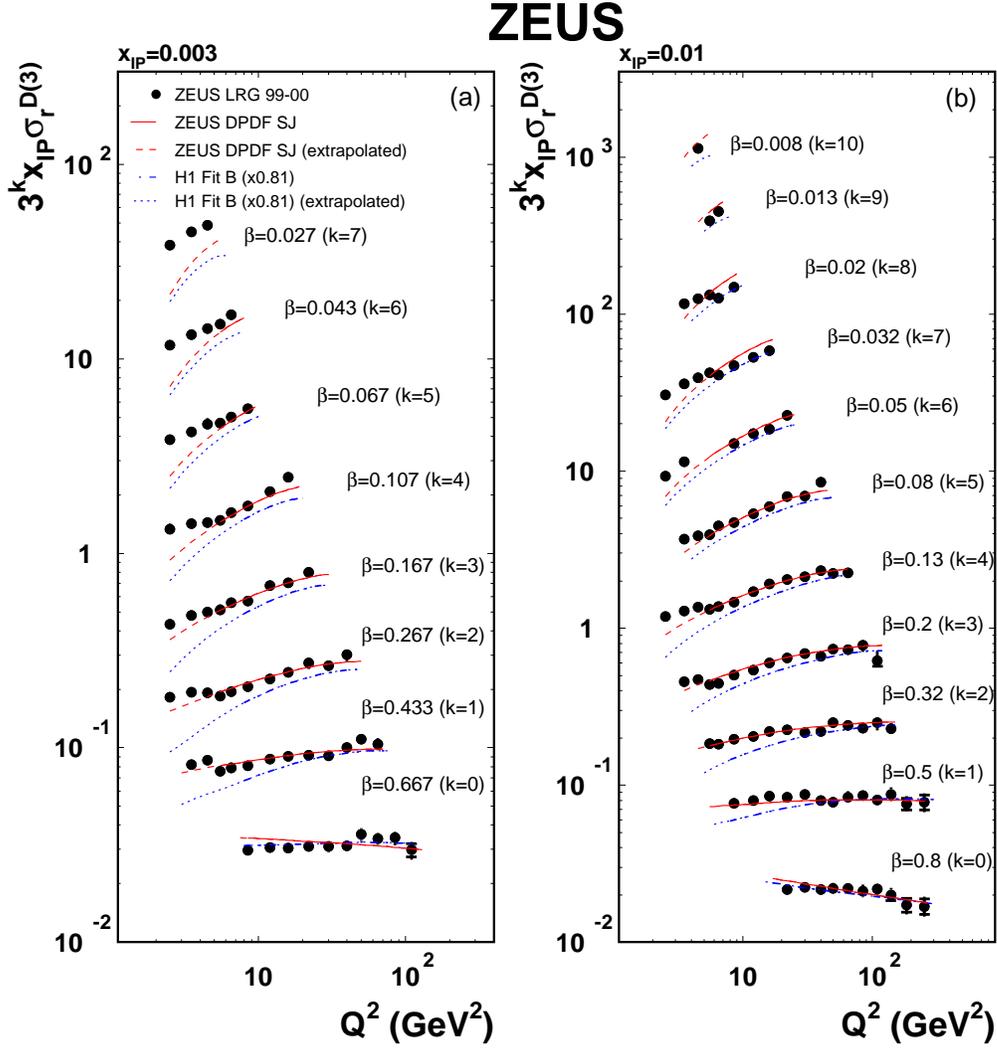}
\end{center}
\caption{The fit \fitJ\ (continuous line) and 
H1 Fit B~\protect\cite{h1-lrg} (dashed-dotted line) compared 
to the ZEUS LRG data~\protect\cite{Chekanov:2008lrg} 
at (a) $x_{\pom}=0.003$ and (b)~$x_{\pom}=0.01$ 
as a function of  $Q^2$ for different $\beta$ 
values. Where visible, the inner error bars show 
the statistical uncertainties and the full bars indicate the 
statistical and systematic uncertainties added in quadrature.
The H1 predictions are corrected to $M_N = m_p$ 
via the scaling factor 0.81. The dashed (dotted) lines 
represent the DGLAP extrapolation beyond the ZEUS (H1) fitted 
region.} 
\label{fig12}
\vfill
\end{figure}

Several DPDF fits are available from the H1 Collaboration, 
among which the most recent are based on an inclusive 
diffractive sample~\cite{h1-lrg} and on a combination 
of data from inclusive and diffractive dijets in 
DIS~\cite{h1-dijets}. 
In Fig.~\ref{fig12}, the diffractive reduced cross  
sections from the ZEUS LRG sample as a function of 
$Q^2$ for fixed $\beta$ and $x_{\pom}$ are compared to 
the fit \fitJ\ as well as to the 
predictions from ``H1 2006 Fit B''~\cite{h1-lrg}. 
The latter, extracted for masses of the dissociative system 
$M_N < 1.6$ GeV, were multiplied by the scaling factor 
0.81~\cite{h1-lrg} to account for the correction to the 
ZEUS elastic case $M_N=m_p$.
For $Q^2 < 5$ GeV$^2$ in the ZEUS case and for $Q^2<8.5$ 
GeV$^2$ in the H1 case, the fits are extrapolated.  
For $\beta < 0.2$,   
the two fits agree in shape throughout the fitted range, 
but the normalisation of the ZEUS curve is above that of H1. 
At higher $\beta$ and where the predictions 
are extrapolated the agreement worsens. 
These features reflect the degree of consistency 
between the ZEUS and H1 data~\cite{martapaul}. 

%----------------------------------------------------------------

\subsection{Comparison to charm and dijet photoproduction data}

\begin{figure}[p]
\vfill
\begin{center}
%\framebox[13.cm]{\rule{0.pt}{10.cm}}
\includegraphics[width=15cm,height=15cm]{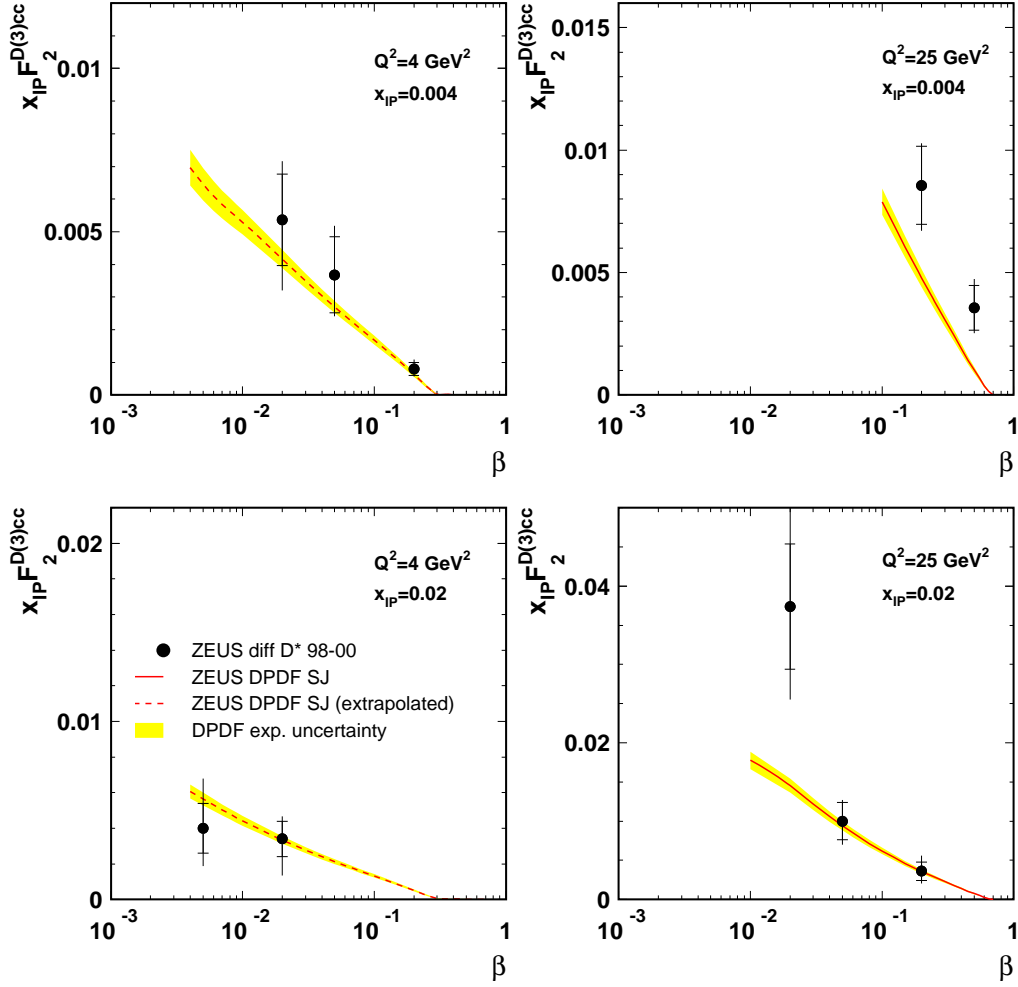}
\end{center}
\caption{\fitJ\ predictions compared to the ZEUS measurement of  
the charm contribution to the diffractive structure function 
multiplied by $x_{\pom}$, $x_{\pom} F_{2}^{D(3)c\overline{c}}$~\protect\cite{Chekanov:2003dstardis} as a function of 
$\beta$ for different $Q^2$ and $x_{\pom}$ values. The inner error bars show 
the statistical uncertainties and the full bars indicate the 
statistical and systematic uncertainties added in quadrature.
The dashed lines represent the extrapolation of the predictions 
beyond the kinematic region where they were obtained. 
The shaded bands show 
the DPDF experimental uncertainty.  
} 
\label{fig13}
\vfill
\end{figure}

Predictions from the fit \fitJ\ are 
compared in Fig.~\ref{fig13} 
to data\cite{Chekanov:2003dstardis} on the charm contribution to the 
diffractive structure function multiplied by $x_{\pom}$, 
$x_{\pom} F_{2}^{D(3)c\overline{c}}$,  
for $x_{\pom}$ of 0.004 and 0.02 and $Q^2$ of 4 and 25 GeV$^2$. 
The predictions are in fair agreement with the data. 
%The fit agrees with the data at low $Q^2$, whereas it undershoots 
%the data for $Q^2$ of 25 GeV$^2$. 

\begin{figure}[p]
\vfill
\begin{center}
%%%\framebox[13.cm]{\rule{0.pt}{10.cm}}
\includegraphics[width=0.6\columnwidth]{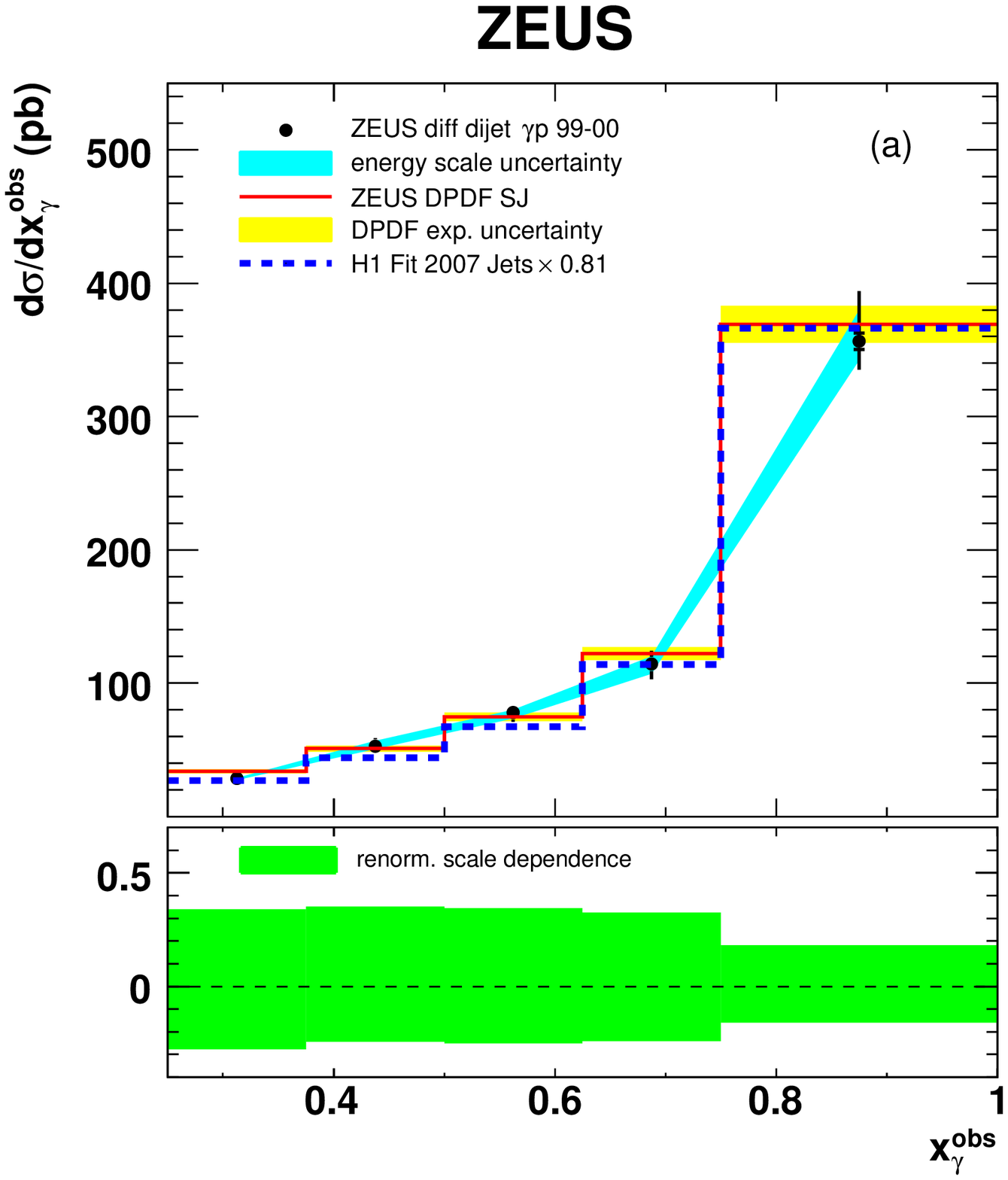}
\includegraphics[width=0.6\columnwidth]{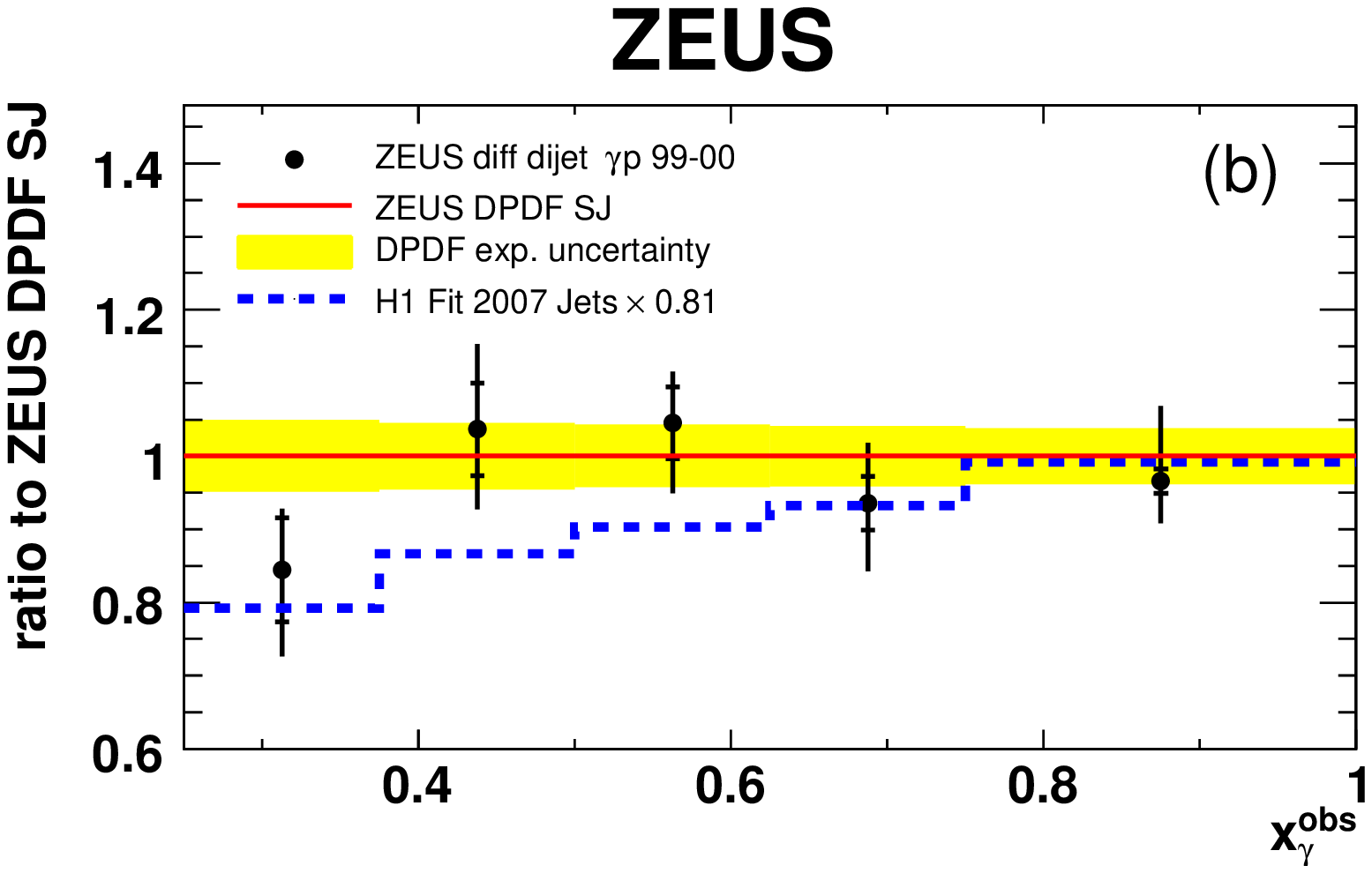}
\end{center}
\caption{(a) \fitJ\ predictions compared to 
the ZEUS diffractive dijet 
photoproduction data~\protect\cite{Chekanov:2007dijetsphp} 
as a function of $x_{\gamma}^{\rm obs}$. The predictions from 
H1 Fit 2007 Jets~\protect\cite{h1-dijets} are also shown, 
corrected to $M_N = m_p$ via the scaling factor 0.81. 
(b) Ratio of the data and of the H1 predictions to the 
ZEUS DPDF SJ predictions. The inner error bars show 
the statistical uncertainties and the full bars indicate the 
statistical and systematic uncertainties added in quadrature. 
The dark shaded bands indicate the jet energy scale uncertainty. 
The light shaded bands show the DPDF experimental uncertainty.
The shaded band at the bottom of (a) shows the renormalisation 
scale uncertainty.
}
\label{fig:php-xg}
\vfill
\end{figure}

\begin{figure}[p]
\vfill
\begin{center}
\includegraphics[width=0.6\columnwidth]{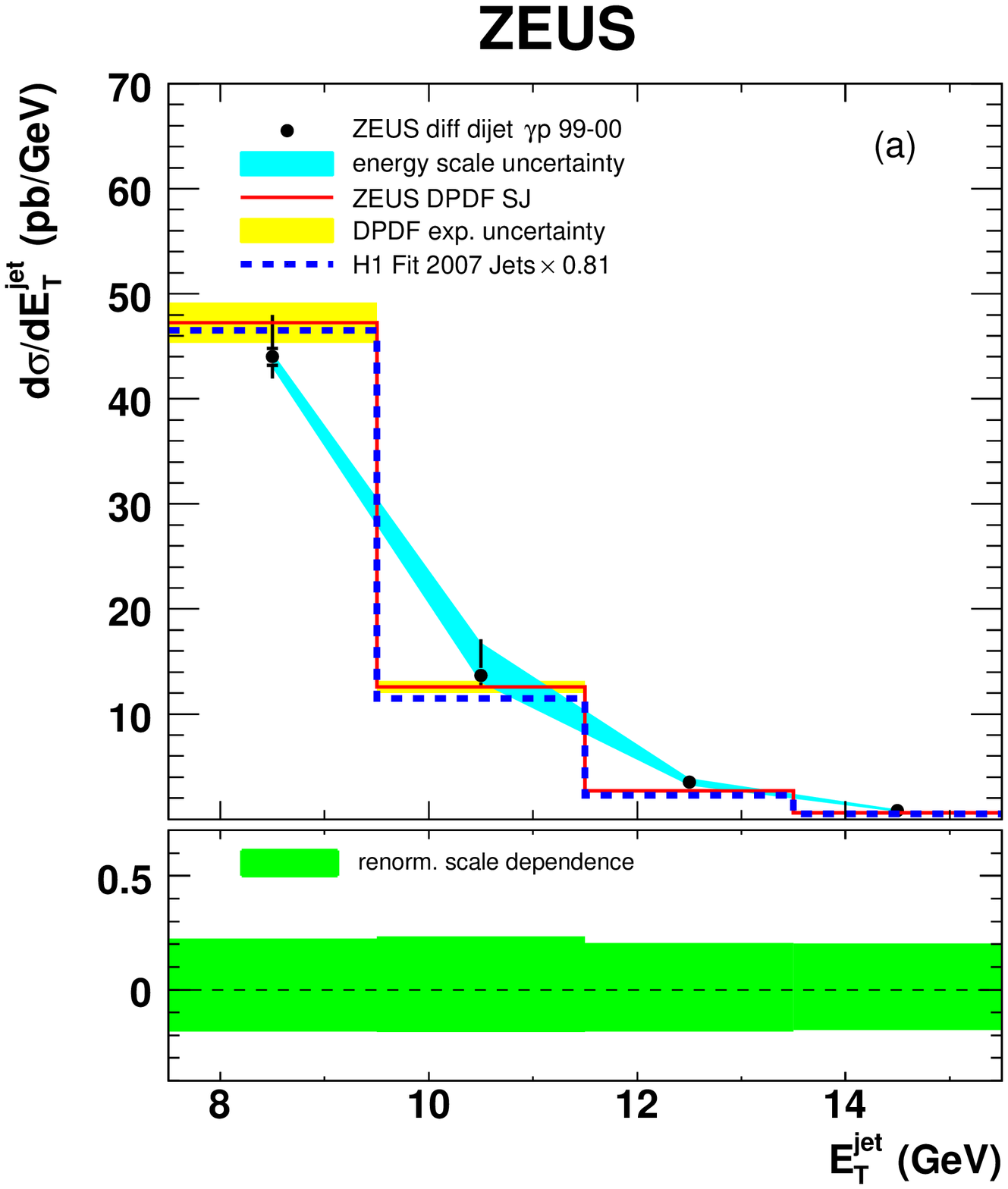}
\includegraphics[width=0.6\columnwidth]{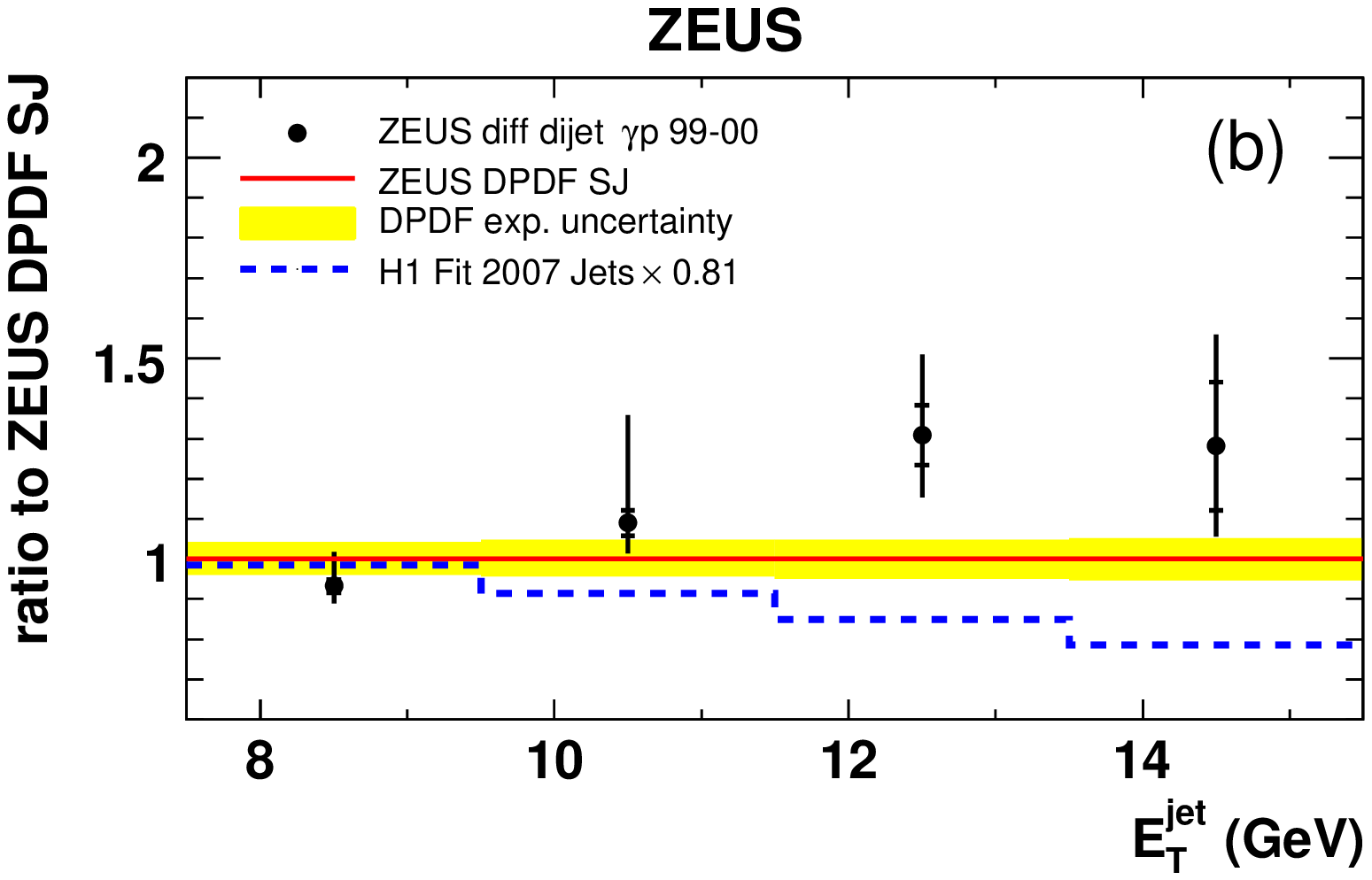}
\end{center}
\caption{(a) \fitJ\ predictions compared to the 
ZEUS diffractive dijet 
photoproduction data~\protect\cite{Chekanov:2007dijetsphp} 
as a function of $\etjet$. The predictions from 
H1 Fit 2007 Jets~\protect\cite{h1-dijets} are also shown, 
corrected to $M_N = m_p$ via the scaling factor 0.81. 
(b) Ratio of the data and of the H1 predictions to 
ZEUS DPDF SJ predictions. Other details as in caption to 
Fig.~\ref{fig:php-xg}.
}
\label{fig:php-et}   
\vfill
\end{figure}

Figures~\ref{fig:php-xg}a and~\ref{fig:php-et}a show the prediction 
from the fit \fitJ\ compared to the diffractive dijet photoproduction 
data~\cite{Chekanov:2007dijetsphp} as a function of $x_{\gamma}^{\rm
  obs}$, the fraction of the photon energy invested in producing 
the dijet
system, and of the tranverse energy $E_T^{\rm jet}$ of the leading   
jet, respectively. 
Dijet photoproduction at leading order in QCD
proceeds through two type of processes: direct processes 
($x_{\gamma}^{\rm obs} \approx$ 1), in which the exchanged photon
interacts as a point-like particle with the partons from the
diffractive exchange; and resolved processes ($x_{\gamma}^{\rm
  obs}<1$), in which the photon behaves as a source of partons
which interact with the diffractive exchange. For the latter, 
as in hadron-hadron interactions, QCD factorisation is not 
expected to hold~\cite{qcdf}. Comparing calculations based on DPDFs extracted 
from other diffractive processes to diffractive dijet photoproduction 
data provides a valuable test of QCD factorisation. The 
predictions shown in Figs.~\ref{fig:php-xg}a and~\ref{fig:php-et}a, obtained 
with the program of Klasen and 
Kramer~\cite{Klasen:1996it}, 
agree adequately with the data over the whole 
$x_{\gamma}^{\rm obs}$ and $E_T^{\rm jet}$ 
ranges\footnote{The program of Frixione and 
Ridolfi~\cite{frixione} was also used and gave the same results.}. 
A further confirmation 
is provided by Figs.~\ref{fig:php-xg}b and \ref{fig:php-et}b, where the ratio 
of the data and the predictions 
is presented as a function of $x_{\gamma}^{\rm obs}$ 
and $E_T^{\rm jet}$, respectively. The ratio is consistent with unity. 
%A systematic rise with $E_T^{\rm jet}$ can't be excluded. 
This reinforces the conclusion of an earlier 
ZEUS publication~\cite{Chekanov:2007dijetsphp}, where the  
data were found to be compatible with no suppression 
either of the resolved component, or of both components
globally. The slight normalisation difference between the predictions
shown here and those of the earlier ZEUS 
publication~\cite{Chekanov:2007dijetsphp} 
is due to the usage of different DPDFs (``H1 2006 Fit B''), 
extracted in a fixed-flavour-number scheme.

Also shown in Figs.~\ref{fig:php-xg} and \ref{fig:php-et} are the 
predictions from ``H1 Fit 2007 Jets''~\cite{h1-dijets}, multiplied 
by the scaling factor 0.81~\cite{h1-lrg} to account for the 
correction to $M_N=m_p$. They overlap with the ZEUS predictions 
in the lowest $E_T^{\rm jet}$ and highest $x_{\gamma}^{\rm obs}$ bins.
At high $E_T^{\rm jet}$ and low $x_{\gamma}^{\rm obs}$, the two predictions 
differ by as much as 20\%. 
%The predictions from the H1 2006 fit B turns out to be higher than 
%the ZEUS fit, which is consistent with the fact that H1 
%DPDFs were extracted in a fixed flavour-number scheme. 

%================================================================
\section{Summary and conclusions}
\label{section:summary}

ZEUS diffractive inclusive cross-section data, together with data 
on diffractive dijet production in DIS, have been used in a NLO
DGLAP QCD analysis to determine the diffractive parton distribution 
functions. Only data with $Q^2 > 5 \,\GeV^2$ could be fitted within the 
combined framework of DGLAP evolution and proton-vertex 
factorisation. 
The extracted DPDFs correspond to 
the single-diffractive reaction with a proton in the final state 
and are valid in the region $|t| < 1$\,GeV$^2$, $M_X > 2$ GeV, 
$x_{\pom} <0.1$. 

NLO QCD predictions based on diffractive parton densities extracted from 
the inclusive data provide a good, simultaneous description of the inclusive 
and dijet data.
The quark densities are well constrained by the inclusive data, whereas 
the sensitivity to the gluon density decreases as the momentum fraction 
$z$ increases. 

The inclusion of the dijet data provides an 
additional constraint on the gluons, allowing the 
determination of both the quark and gluon densities with 
good accuracy. 

Predictions based on the extracted parton densities 
give a fair description of diffractive charm production data; 
also, they are in good agreement with diffractive dijet photoproduction 
cross sections over the whole kinematic region, thus indicating 
no suppression of the resolved component.

\newpage

\vspace{0.5cm}
\noindent {\Large\bf Acknowledgements}
\vspace{0.3cm}

The design, construction and installation of the ZEUS detector were made possible 
by the ingenuity and dedicated efforts of many people inside DESY and from the 
home institutes who are not listed as authors. Their contributions are acknowledged 
with great appreciation. We thank the DESY directorate for their strong support and 
encouragement, and the HERA machine group for their diligent effort. We are grateful for the support of the DESY computing and network services.

%%% Local Variables: 
%%% mode: latex
%%% TeX-master: t
%%% End: 

%------------------------------------------------------------------------------
%       Bibliography
%------------------------------------------------------------------------------
\providecommand{\etal}{et al.\xspace}
\providecommand{\coll}{Coll.\xspace}
\catcode`\@=11
\def\@bibitem#1{%
\ifmc@bstsupport
  \mc@iftail{#1}%
    {;\newline\ignorespaces}%
    {\ifmc@first\else.\fi\orig@bibitem{#1}}
  \mc@firstfalse
\else
  \mc@iftail{#1}%
    {\ignorespaces}%
    {\orig@bibitem{#1}}%
\fi}%
\catcode`\@=12

%------------------------------------------------------------------------------
%       Tables
%------------------------------------------------------------------------------
%%%\documentclass[12pt]{article}
 
%%%\begin{document}
%%%\newcommand {\pom}  {I\hspace{-0.2em}P} 
%%%\newcommand {\xpom} {\mbox{$x_{_{\pom}}$}} 

\begin{table}[ht]
\begin{center}
\begin{tabular}{|c|c|c|c|c|}
\hline
Parameter & Fixed to & Measurement & Ref.\\
\hline
$\alpha'_{\pom}  $ & $0$ & $-0.01 \pm 0.06 ({\rm stat.})  
^{+0.04}_{-0.08} ({\rm syst.}) \pm 0.04 ({\rm model})$\,GeV$^{-2}
$ & ~\cite{Chekanov:2008lrg}\\
$\alpha'_{\reg}$ & $0.9$\,GeV$^{-2}$ & $0.90 \pm 0.10 $\,GeV$^{-2}$ & ~\cite{softpom}\\
$B_{\pom}$ & 7.0\,GeV$^{-2}$ & $7.1 \pm 0.7 ({\rm stat.})^{+1.4}_{-0.7}({\rm syst.})$\,GeV$^{-2}$ & ~\cite{Chekanov:2008lrg}\\
$B_{\reg}$ & 2.0\,GeV$^{-2}$ &$2.0 \pm 2.0$\,GeV$^{-2}$ & ~\cite{softpom}\\
\hline
\end{tabular}
\caption{The values of the parameters fixed in the fits 
and the measurements providing this input.}
\label{tab-pomeronparam}
\end{center}
\end{table}

\begin{table}[ht]
\begin{center}
\begin{tabular}{|l|l|l|}
\hline
Fit name & Data set & $zg(z)$\\
\hline
\fitS & LRG + LPS & $A_g\, z^{B_g} (1-z)^{C_g}$ \\
\fitC & LRG + LPS & $A_g$\\
\fitJ & LRG + LPS + DIS dijets & $A_g\, z^{B_g} (1-z)^{C_g}$ \\
\hline
\end{tabular}
\caption{Data sets used and parameterisation of the gluon density at the starting scale for the different fits.}
\label{tab-fit}
\end{center}
\end{table}

\begin{table}[ht]
\begin{center}
\begin{tabular}{|c|rcl|rcl|rcl|}
\hline
Parameter & 
\multicolumn{3}{c|}{ Fit value} & 
\multicolumn{3}{c|}{ Fit value} &
\multicolumn{3}{c|}{ Fit value}\\
& 
\multicolumn{3}{c|}{DPDF S} & 
\multicolumn{3}{c|}{DPDF C} &
\multicolumn{3}{c|}{DPDF SJ} \\
\hline
$A_q$ & 0.135 & $\pm$ & 0.025 & 0.161 & $\pm$ & 0.030 & 0.151 & $\pm$ & 0.020\\
$B_q$ & 1.34  & $\pm$ & 0.05 & 1.25  & $\pm$ & 0.03 & 1.23  & $\pm$ & 0.04\\
$C_q$ & 0.340 & $\pm$ & 0.043 & 0.358 & $\pm$ & 0.043 & 0.332 & $\pm$ & 0.049\\
$A_g$ & 0.131 & $\pm$ & 0.035 & 0.434 & $\pm$ & 0.074 & 0.301 & $\pm$ & 0.025\\
$B_g$ & $-0.422$ & $\pm$ & $0.066$ & & $0$ & & $-0.161$ & $\pm$ & $0.051$ \\
$C_g$ & $-0.725$ & $\pm$ & $0.082$ & & $0$ & & $-0.232$ & $\pm$ & $0.058$ \\
$\alpha_{\pom}(0)$ & 1.12 & $\pm$ & 0.02 & 1.11 & $\pm$ & 0.02 & 1.11 & $\pm$ & 0.02\\
$\alpha_{\reg}(0)$ & 0.732 & $\pm$ & 0.031 & 0.668 & $\pm$ & 0.040 & 0.699 & $\pm$ & 0.043\\
$A_{\reg}$ & 2.50 & $\pm$ & 0.52 & 3.41 & $\pm$ & 1.27 & 2.70 & $\pm$ & 0.66\\
%% $\chi^2$/ndf & 314.9/265 =  & 311.7/265 & 335.77/293\\
$\chi^2$/ndf & 315/265 & = & 1.19 & 312/265 & = & 1.18 & 336/293 & = & 1.15\\
\hline
\end{tabular}
\caption{Parameters obtained with the different fits and their 
experimental uncertainties.}
\label{tab-pdf}
\end{center}
\end{table}

%%%%%%\end{document}  

%------------------------------------------------------------------------------
%       Figures
%------------------------------------------------------------------------------
%---------------------------------------------
%
%---------------------------------------------

%---------------------------------------------

%---------------------------------------------

%---------------------------------------------

%---------------------------------------------

%---------------------------------------------

%---------------------------------------------

%---------------------------------------------

%---------------------------------------------

%---------------------------------------------

%---------------------------------------------

%---------------------------------------------

%---------------------------------------------

\clearpage

%
%       ... that's it
%

\begin{mcbibliography}{10}

\bibitem{h1-lrg}
H1 Coll., A. Aktas et al.,
\newblock Eur. Phys. J.{} {\bf C~48},~715~(2006)\relax
\relax
\bibitem{h1-dijets}
H1 \coll, A.~Aktas \etal,
\newblock JHEP{} {\bf 0710:042}~(2007)\relax
\relax
\bibitem{MRW2007}
A.D. Martin, M.G. Ryskin and G. Watt,
\newblock Phys.\ Lett.{} {\bf B 644},~131~(2007)\relax
\relax
\bibitem{GB2008}
K. Golec-Biernat and A. Luszczak,
\newblock arXiv:0812.3090 (2008){}\relax
\relax
\bibitem{dglap1}
V.N.~Gribov and L.N.~Lipatov,
\newblock Sov. J. Nucl. Phys.{} {\bf 15},~438~(1972)\relax
\relax
\bibitem{dglap2}
Yu.L.~Dokshitzer,
\newblock Sov. Phys. JETP{} {\bf 46},~641~(1977)\relax
\relax
\bibitem{dglap3}
G.~Altarelli and G.~Parisi,
\newblock Nucl. Phys.{} {\bf B~126},~298~(1977)\relax
\relax
\bibitem{Chekanov:2005nn}
ZEUS \coll, S.~Chekanov \etal,
\newblock Eur. Phys. J.{} {\bf C42},~1~(2005)\relax
\relax
\bibitem{Chekanov:2008lrg}
ZEUS \coll, S.~Chekanov \etal,
\newblock Nucl. Phys.{} {\bf B~816},~1~(2009)\relax
\relax
\bibitem{Chekanov:2007dijetsdis}
ZEUS \coll, S.~Chekanov \etal,
\newblock Eur. Phys. J.{} {\bf C~52},~813~(2007)\relax
\relax
\bibitem{Chekanov:2003dstardis}
ZEUS \coll, S.~Chekanov \etal,
\newblock Nucl. Phys.{} {\bf B 672},~3~(2003)\relax
\relax
\bibitem{Chekanov:2007dijetsphp}
ZEUS \coll, S.~Chekanov \etal,
\newblock Eur. Phys. J.{} {\bf C~55},~177~(2008)\relax
\relax
\bibitem{qcdf}
J.C.~Collins,
\newblock Phys. Rev.{} {\bf D~57},~3051~(1998)\relax
\relax
\bibitem{qcdf1}
Erratum,
\newblock ibid.{} {\bf D~61},~019902~(2000)\relax
\relax
\bibitem{Trentadue}
L. Trentadue and G. Veneziano,
\newblock Phys. Lett.{} {\bf B~323},~201~(1994)\relax
\relax
\bibitem{Berera}
A. Berera and D.E. Soper,
\newblock Phys. Rev.{} {\bf D~53},~6162~(1996)\relax
\relax
\bibitem{regge}
P.D.B.~Collins,
\newblock {\em An Introduction to {Regge} Theory and High Energy Physics}.
\newblock Cambridge University Press, Cambridge (1997)\relax
\relax
\bibitem{Chekanov:2002nlo}
ZEUS \coll, S.~Chekanov \etal,
\newblock Phys. Rev.{} {\bf D~67},~012007~(2003)\relax
\relax
\bibitem{f2bpc}
ZEUS \coll, J.~Breitweg \etal,
\newblock Phys. Lett.{} {\bf B~407},~432~(1997)\relax
\relax
\bibitem{tr2}
R.S. Thorne and R.G. Roberts,
\newblock Phys. Rev.{} {\bf D~57},~6871~(1998)\relax
\relax
\bibitem{disent}
S. Catani and M.H.~Seymour,
\newblock Nucl. Phys.{} {\bf B~510},~503~(1998)\relax
\relax
\bibitem{nlojet}
Z. Nagy and Z. Trocsanyi,
\newblock Phys. Rev. Lett. {\bf 87} (2001) 082001{}\relax
\relax
\bibitem{tesi_bonato}
A.~Bonato, Ph. D. Thesis, Hamburg University, DESY-THESIS-2008-008\relax
\relax
\bibitem{grv1}
M. Gl\"{u}ck, E. Reya and A. Vogt,
\newblock Z. Phys.{} {\bf C~53},~127~(1992)\relax
\relax
\bibitem{mx1}
ZEUS Coll., S. Chekanov \etal,
\newblock Nucl. Phys.{} {\bf B~713},~3~(2005)\relax
\relax
\bibitem{mx2}
ZEUS \coll, S.~Chekanov \etal,
\newblock Nucl. Phys.{} {\bf B 800},~1~(2008)\relax
\relax
\bibitem{lps+charm}
ZEUS \coll, S.~Chekanov \etal,
\newblock Eur. Phys. J.{} {\bf C38},~43~(2004)\relax
\relax
\bibitem{martapaul}
P. Newman and M. Ruspa,
\newblock {\em Proc. of the Workshop HERA and the LHC - 2nd Workshop on the
  implications of HERA for the LHC physics}.
\newblock H. Jung and A. De Roeck eds., Hamburg (2009),
  DESY-PROC-2009-002\relax
\relax
\bibitem{Klasen:1996it}
M. Klasen and G. Kramer,
\newblock Z. Phys.{} {\bf C76},~67~(1997)\relax
\relax
\bibitem{frixione}
S. Frixione and G. Ridolfi,
\newblock Nucl. Phys.{} {\bf B~507},~315~(1997)\relax
\relax
\bibitem{softpom}
A.~Donnachie and P.L.~Landshoff,
\newblock Phys. Lett.{} {\bf B~296},~227~(1992)\relax
\relax
\end{mcbibliography}
\end{document}